\documentclass[aps,pre,floatfix,nofootinbib,showpacs]{revtex4} 
\usepackage{graphicx} \usepackage{amsmath} \usepackage{amssymb} 
\usepackage[sort&compress]{natbib}
\usepackage{graphicx}
\usepackage{amsmath}
\usepackage{amssymb}
\newcommand{\BEQ}{\begin{equation}}
\newcommand{\EEQ}{\end{equation}}
\newcommand{\BEA}{\begin{eqnarray}}
\newcommand{\EEA}{\end{eqnarray}}
\newcommand{\nn}{\nonumber \\}
\renewcommand{\d}{{\rm d}}

\renewcommand{\nn}{\nonumber}
\newcommand{\lel}{\left\langle}
\newcommand{\rir}{\right\rangle}
\newcommand{\eps}{\varepsilon}
\newcommand{\si}{\hat{\sigma}}
\newcommand{\siz}{\hat{\sigma}_z}
\newcommand{\six}{\hat{\sigma}_x}
\newcommand{\siy}{\hat{\sigma}_y}
\newcommand{\sip}{\hat{\sigma}_{+}}
\newcommand{\oneh}{\frac{1}{\hbar}}
\newcommand{\simin}{\hat{\sigma}_{-}}
\newcommand{\sipm}{\hat{\sigma}_{\pm}}

\newcommand{\om}{\omega}

\newcommand{\e}{\hat{\eta }}
\newcommand{\ha}{\hat{a}}
\newcommand{\hb}{\hat{b}}
\newcommand{\X}{\hat{X}}
\newcommand{\U}{\hat{U}}
\newcommand{\taut}{\delta}

\newcommand{\half}{\frac{1}{2}}

\newcommand{\CE}{{\cal E}} 
\newcommand{\CP}{{\cal P}}

\newcommand{\piplus}{\hat{\Pi}_{+}}
\newcommand{\piminus}{\hat{\Pi}_{-}}
\newcommand{\pipm}{\hat{\Pi}_{\pm}}
\newcommand{\pimp}{\hat{\Pi}_{\mp}}
\newcommand{\HH}{\hat{H}}
\newcommand{\HF}{\hat{H}_{\rm F}}
\newcommand{\HS}{\hat{H}_{\rm S}}
\newcommand{\HI}{\hat{H}_{\rm I}}
\newcommand{\HB}{\hat{H}_{\rm B}}
\newcommand{\tr}{{\rm tr}}
\renewcommand{\thesection}{\arabic{section}}

\def\dbarrm {{\mathchar'26\mkern-11mu{\rm d}}}                       %
\begin{document} 
\draft
\title
{Work extraction in the spin-boson model.}
\date{\today}
\author{A.E. Allahverdyan$^{1,2)}$,
R. Serral Graci\`a$^{1)}$
and Th.M. Nieuwenhuizen$^{1)}$}
\address{
$^{1)}$ Institute for Theoretical Physics,
University of Amsterdam,
Valckenierstraat 65, 1018 XE Amsterdam, The Netherlands,\\
$^{2)}$Yerevan Physics Institute,
Alikhanian Brothers St. 2, Yerevan 375036, Armenia
}

\begin{abstract}

We show that work can be extracted from a two-level system (spin)
coupled to a bosonic thermal bath. This is possible due to different
initial temperatures of the spin and the bath, both positive (no spin
population inversion) and is realized by means of a suitable sequence
of sharp pulses applied to the spin. The extracted work can be of the
order of the response energy of the bath, therefore much larger than
the energy of the spin. Moreover, the efficiency of extraction can be
very close to its maximum, given by the Carnot bound, at the same time
the overall amount of the extracted work is maximal. Therefore, we get
a finite power at efficiency close to the Carnot bound.

The effect comes from the backreaction of the spin on the bath, and it
survives for a strongly disordered (inhomogeneously broadened)
ensemble of spins. It is connected with generation of coherences during
the work-extraction process, and we derived it in an exactly solvable
model.  All the necessary general thermodynamical relations are
derived from the first principles of quantum mechanics and connections
are made with processes of lasing without inversion and with quantum heat
engines.

\end{abstract}

\maketitle


\renewcommand{\thesection}{\arabic{section}}
\section{Introduction.}
\setcounter{equation}{0}\setcounter{figure}{0} 
\renewcommand{\thesection}{\arabic{section}.}

A known feature of technological progress is the increase of human
ability to control and design the microscopic world.  Recent efforts
in manipulating simple quantum systems e.g., in the context of quantum
computing or quantum chemistry, is one aspect of this general
trend. Another aspect is the field of quantum thermodynamics whose
main objective is in designing and studying new thermodynamical
processes in the domain where quantum features of matter are
relevant. In particular, this activity aims to improve our
understanding of the standard thermodynamics \cite{landau,klim,balian}
by addressing its concepts from the first principles of quantum
mechanics \cite{ANprl,NAlinw,gemmer}.  The current activity in quantum
thermodynamics includes quantum engines
\cite{alicki,kosloff,bender,marlan,chen1,linke}, general aspects of work
extraction from quantum systems \cite{abnmaximalwork}, thermodynamical
aspects of quantum information theory \cite{lloyd,horodecki}, and
limits of thermodynamical concepts such as the second law
\cite{ANprl,ANjpa,NAlinw} and the temperature \cite{michel}.  There were
also much earlier applications concerning, in particular,
thermodynamic aspects of lasers and masers \cite{lasers}.

Our present purpose is to study work extraction from a two-temperature
system on the basis of the known spin-boson model
\cite{petr,rmp,lu,viola}: a two-level system coupled to a bosonic
thermal bath. The motivation to use a two-level system is nearly
obvious, it is almost everywhere, and it is the minimal model having
non-trivial quantum features. The necessity of the bath has to be
stressed separately since, in the usual practice of quantum systems
manipulation, the bath is a serious hindrance.  As follows, the process
of work-extraction really needs external thermal baths: The second law
in Thomson's formulation |which is derived as a theorem in quantum
mechanics \cite{thirring,bk,bassett,woron,lenard,ANthomson}| forbids
work-extraction from an equilibrium system by means of cyclic processes
generated by external fields. The easiest way to employ an equilibrium
system in work-extraction is to attach it to a thermal bath having a
different temperature, thus forming a local-equilibrium state. The
overall system is then out of equilibrium and work-extraction from
cycles is not forbidden, at least in principle. This was shown
explicitly in \cite{ANprl,ANjpa,NAlinw}. 

This general restriction determined the way how standard quantum
work-extraction (also known as amplification or lasing/masing)
processes are designed \cite{lasers}.  The most traditional lasers and
masers operate by extracting work from an ensemble of two-level
systems having a negative temperature, in other words, population
inversion, which is a strongly non-equlibrium state.  More recent
schemes of lasing without inversion employ non-equilibrium states of
three (four, multi) level systems without population inversion of
energy levels, but with initially sizable
non-diagonal terms of the corresponding density matrix in the energy
representation, usually called coherences \cite{opt,olga}. These
schemes attracted attention due to both their conceptual novelty and
the fact that non-zero non-diagononal elements represent a weaker form
of non-equilibrium than population inversion, and thus their
preparation can be an easier task \cite{opt,olga}.

The mechanism of work-extraction proposed in the present paper differs
from the standard ones in several aspects: 
\begin{itemize}

\item Work-extraction
(amplification, lasing) can be achieved in two-level systems without
population difference and without using an initially coherent state. A
setup consisting of a positive temperature spin interacting with a
thermal bath at some higher or lower temperature suffices to extract
work and thereby amplify pulsed fields acting on the spin.  Moreover,
the extracted work can be of the order of the bath's response energy,
which is larger than the energy of the spin.  Thus, when viewed
as lasing without inversion, the presented mechanism offers definite
advantages over the existing schemes.

\item
The effect survives for a disordered ensemble of spins, where the
spin have a random energy with a large dispersion. The reason of the
survival is the possibility to combine the work-extraction process
with the spin-echo phenomena \cite{Hahn,Waugh}.  As a consequence, we
have a phenomenon even more amazing than the original spin-echo: 
a high-temperature, completely disordered ensemble of spins
can serve as a medium of work-extraction.

\item The efficiency of work-extraction can approach its maximally
possible value given by Carnot bound. Moreover, the efficiency is
maximized simultaneously with the overall amount of the extracted work.
In addition, the power of work (i.e. the work divided over the total
duration of the work-extraction process) is finite. Thus, in marked
contrast to the original Carnot cycle \cite{landau,balian} and some of
its realizations in quantum engines \cite{linke}, the three basic
objectives of a good work-extraction process are met together: large
amount of extracted work, high efficiency, and finite power. 

\end{itemize}

The origin of the presented mechanism is that, besides well-known
effects of dissipation and decoherence induced by a thermal bath on a
spin interacting with it, there is another effect the presence of which is
frequently not acknowledged. This is the backreaction of the spin to
the bath, which in combination with external fields influences the
spin's dynamics. The effect exists even for relatively small |but
generic| bath-spin coupling constants, but is typically neglected from
standard weak-coupling theories \cite{lindblad}. Our present treatment
of the bath-spin interaction is exact and allows to study the full
influence of backreaction and memory effects.

This paper is organized as follows. In section \ref{themodel} we
recall a version of the spin-boson model we work with. It nowdays
became one of the most popular models in the theory of open quantum
systems \cite{petr,rmp,lu,viola,lidar}.
Section \ref{pulseddynamics} describes the action of
external fields and discusses the definition of work.
The next section presents experimental
realizations of our setup in various situations of two-level system(s)
interacting with a thermal bath. Section
\ref{work-extraction} discusses general limits of work-extraction
from a two-temperature system. The next two sections describe our
basic results on the work-extraction, effciency, and the power of
work.  The last section offers our main conclusions and compares our
results with the ones existing in literature. Several technical
questions are considered in Appendices.  We have tried to make this
paper reasonably self-contained.  This especially concerns the
concepts and relations of the standard thermodynamics, which are not
accepted uncritically, but, in many situations, are derived from the
first principles of quantum mechanics.

\renewcommand{\thesection}{\arabic{section}}
\section{ The model}
\setcounter{equation}{0}\setcounter{figure}{0} 
\renewcommand{\thesection}{\arabic{section}.}
\label{themodel}

As common when dealing with open systems, 
the Hamiltonian $\HH$ is composed by three parts:
\BEA
\label{ham}
\HH=\HS+\HB+\HI.
\EEA
$\HS$ stands for the Hamiltonian of a two level system (spin $\half$):
\BEA
\label{hamspin}
\HS=\frac{\eps}{2}\,\si _z,\qquad \eps\equiv\hbar\Omega,
\EEA
where $\si _x$, $\si _y$ and $\si _z$ are Pauli's matrices, 
and where the energy levels are $\pm\frac{\eps}{2}$. 

The spin interacts with a thermal bath which is a set of
harmonic oscillators. In some cases this may be taken in the literal
sense, when harmonic oscillators represent phonons or photons. It is
also known that rather general classes of thermal baths can be
effectively represented via harmonic oscillators ~\cite{ms,cl}. 
Thus for the Hamiltonian of the bath we take:
\BEA
\label{hambath}
\HB=\sum _k\hbar\om _k\ha^{\dagger}_k\ha _k, 
\qquad [\ha _l,\ha ^{\dagger}_k]=\delta _{kl},
\EEA where
$\ha_k^{\dagger}$ and $\ha _k$ are creation and annihilation operators of
the bath oscillator with the index $k$. The thermodynamic limit for the
bath will be taken later on.

The next important point is to specify the interaction between the
spin and the bath. Recall that any reasonable model of a thermal bath
is expected to drive a non-stationary state of the spin towards a
stationary state. In this respect, for two-level systems, one
distinguishes two types of relaxation processes and the
corresponding times scales \cite{balian,lasers,slichter,nmr,abo}:

\begin{enumerate}
\item
${\cal T}_2$-time scale related to the relaxation of the average
transversal components $\langle\six\rangle$ and $\langle\siy\rangle$
of the spin (decoherence).  Note that the very notion of the
transversal components is defined by the form (\ref{hamspin}) of the
spin Hamiltonian.

\item ${\cal T}_1$-time scale related to the relaxation of
$\langle\siz\rangle$.  It is customary to have situations, where
\BEA
\label{tristan}
{\cal T}_2\ll {\cal T}_1,
\EEA
the main physical reason being that the transversal components are not
directly related to the energy of the spin. 
\end{enumerate}
Our basic assumption on the relaxation times is (\ref{tristan})
\footnote{There is also a third relaxation time ${\cal T}_2^*$ which has a 
different origin. It only appears when dealing with an ensemble of
non-interacting spins each having Hamiltonian (\ref{hamspin}) with a
randomly distributed energy $\eps$ (dephasing). The
influence of ${\cal T}_2^*$ is studied in section \ref{spinecho}.}. 
Moreover, to
facilitate the solution of the model we will disregard ${\cal T}_1$
time as being very large, thereby restricting the times of our
interest to those much shorter than ${\cal T}_1$. The interaction Hamiltonian
is thus chosen such that it induces only transversal relaxation:
\BEA
\label{hamint}
\HI=\frac{\hbar}{2}\X\si _z,\qquad
\X\equiv\sum _kg_k(\ha _k^{\dagger}+\ha _k),
\EEA
where $g_k$ are the coupling constants to be specified later, 
and where $\X$ is the collective coordinate operator of the bath.

The last ingredient of our model is external fields which are acting
on the spin. However, before discussing them in the next section, we
shall recall how the model with Hamiltonian Eq. (\ref{ham}) is solved
without external fields. 

\subsection{Heisenberg equations and their exact solution.}

Heisenberg equations for operators $\siz(t)$ and $\ha_k (t)$ read from
(\ref{hamspin}, \ref{hambath},
\ref{hamint}, \ref{ham}):
\BEA
\label{komnin}
\dot{\si}_z=0,\qquad \siz(t)=\siz(0),
\EEA
\BEA
\label{besh}
\dot{\ha } _k=\frac{i}{\hbar}[H,\ha _k] 
=-i\om _k\ha _k-\frac{i}{2}g_k\si _z.
\EEA
Eqs.~(\ref{komnin}, \ref{besh}) are solved as 
\BEA
\label{gegel}\label{akt=}
\ha _k(t)=
e^{-i\om _kt}\ha _k(0)+\frac{g_k\si _z}{2\om _k}\,\left(
e^{-i\om _kt}-1
\right),
\EEA
and then
\BEA
\label{kant}
\X(t)=\e (t)-\si_z\,G(t),
\EEA
where
\BEA
G(t)\equiv
\sum _k \frac{g^2_k}{\om _k}(1-\cos \om _kt),
\label{transilvan}
\EEA
quantifies the reaction of the spin on the collective operator of the bath,
and where we denoted
\BEA \label{eta=}
\e  (t)=
\sum_kg_k[\ha _k^{\dagger}(0)e^{i\om _kt}+\ha _k(0)e^{-i\om _kt}],
\EEA
for the quantum noise operator \footnote{\label{es1}
Note that the commutator of the
quantum noise is a c-number:
$[\,\e (t),\e (s)]={-2}{i}\,\sum_kg_k^2\,\sin \om_k(t-s)
=-2i\hbar\dot{G}(t-s)
$.}. Recalling the standard relations
\BEA
\sipm =\si _x\pm i\,\si _y,\qquad
[\siz,\sipm]=\pm 2\sipm,\qquad \si _z\sipm =\pm\sipm, 
\EEA
and using (\ref{kant}) and $[\X(t),\sipm(t)]=0$ |since they belong to
different Hilbert spaces,| one derives
\BEA
\label{k3}\label{k1}
\dot{\si}_\pm=\frac{i}{\hbar}[H,\sipm] =\pm i
\left(\,\Omega +\X\right)\,\sip=i
\,\left(\pm\Omega \pm\e(t)-G(t)\,\right)\,\sipm.
\EEA
These equations are solved as:
\BEA
\label{k4}
&&\hat{\sigma}_\pm (t)=\exp \left [\pm\, i\Omega t-iF(t)
\right]\,\hat\Pi_\pm (0,t)\,\hat{\sigma}_\pm (0),\\
&&\hat{\Pi}_{\pm}(t_0,t_1)\equiv\,{\cal T}\,\exp\left[\pm
i\int_{t_0}^{t_1}\,\d s\,\e (s)
\right],\label{pi=}\\
&&F(t)\equiv
\int_0^t\,\d s\, G(s)=\sum _k \frac{g^2_k}{\om _k}\left
(t-\frac{\sin \om _kt}{\om_k}\right),
\EEA
where ${\cal T}$ stands for the time-ordering operator.
It is seen again from (\ref{k4}) that there are two effects 
generated by the bath-spin interaction:
besides random influences entering with
the quantum noise $\e(t)$, there is a deterministic influence
generated by the backreaction term $F(t)$, 
somewhat similar to damping (friction) in the problem of quantum
brownian motion.

\subsection{Factorized initial conditions.}

Let us assume that initially, at the moment
$t=0$, the bath and the spin are in the following factorized state:
\BEA
\label{fedor}
\rho (0)=\rho _{\rm S}(0)\otimes \rho _{\rm B}(0)=\rho _{\rm S}(0)\otimes
\frac{e^{-\beta \HB}}{\tr\,e^{-\beta \HB}}
\EEA
where $\rho_{\rm S}(0)$ is the initial density matrix of the spin, and
where the bath is initially at equilibrium with temperature $T=1/\beta$. 

Factorized initial conditions are adequate when the spin is prepared
independently from the equilibrium bath and then is brought in contact
to it at the initial time
\footnote{It is useful to note that this process of bringing spin
in contact to the bath need by itself not be connected with any
fundamental energy cost. Imagine, fo example, a sudden switching of
the interaction Hamiltonian $\HI=\frac{1}{2}\X\siz$. Since in the
equilibrium state of the bath $\langle \X\rangle=0$, the work done for
the realization of this switching is zero.}.  For example, injection
of an electronic spin into a quantum dot, or creation of an exciton by
external radiation.  Yet another situation where factorized initial
conditions can be adopted is a (strong) selective measurement of
$\siz$ by an external apparatus. In this case $\rho _{\rm S}(0)$ is an
eigenstate of $\siz$ upon which the selection was done.
Non-factorized initial states are commented upon below, in section
\ref{corini}.

The equilibrium relation 
\BEA
&&\langle \ha _k^\dagger (0)\rangle=
\langle\ha _k(0)\rangle =0,\\
&&\langle \ha _k^\dagger (0) \ha _k(0)+\ha _k(0)\ha _k^\dagger(0)\rangle
=\coth \left (\frac{\beta\hbar\omega_k}{2}\right)
\EEA
derived from (\ref{fedor}),
imply that the quantum noise is a stationary Gaussian operator with 
\BEA
\langle \e (t)\rangle =0,
\EEA
and having the time-ordered correlation function:
\BEA
\label{02}
K_{\cal T}(t-t')=\langle {\cal T}\left [\e  (t) \e  (t')\right ]
\rangle _{\e }=
\sum_k\,g^2_k\left [\, \coth \left
(\frac{\beta\hbar\omega_k}{2}\right)\, \cos \om_k (t-t') 
-i\,{\rm sgn}\, (t-t')\sin \om_k (t-t')
\right]
\EEA
where the average $\langle ...\rangle $ is taken over 
the initial state (\ref{fedor}). It can be written as
\BEA
K_{\cal T}(t)=K(t)-i\,\dot G(t),
\EEA
where
\BEA
\label{03}
K(t-t')=\Re K_{\cal T}(t-t') 
=\half \langle \e  (t) \e  (t')+\e  (t') \e  (t)
\rangle=
\sum_k\,g^2_k\, \coth \left
(\frac{\beta\hbar\omega_k}{2}\right)\, \cos \om_k (t-t'),
\EEA
is the symmetrized correlation function.

Since $\e (t)$ is a gaussian random operator, one can use Wick's
theorem for decomposing higher-order products 
\footnote{Recalling Wick's theorem (for $\langle \eta\rangle=0$):
Any correlation of an odd number of $\e$'s vanishes. A
correlation of an even number of $\e$'s is equal to the sum of
products of pair correlations, the sum being taken over all pairings.
For example:
\BEA
\label{wick}
\langle{\cal T}\, \e (t_1)\e (t_2)\e (t_3)\e (t_4)\rangle =
\langle{\cal T}\, \e (t_1)\e (t_2)\rangle 
\langle{\cal T}\, \e (t_3)\e (t_4)\rangle 
+\langle{\cal T}\,\e (t_1)\e (t_3)\rangle 
\langle{\cal T}\, \e (t_2)\e (t_4)\rangle 
+\langle{\cal T}\, \e (t_1)\e (t_4)\rangle 
\langle{\cal T}\, \e (t_2)\e (t_3)\rangle.\nonumber
\EEA
Note that the similar Wick-decomposition of 
$\langle{\cal T}\, \e (t_1)...\e (t_{2k})\rangle$ will be a sum of
$(2k-1)!!=(2k-1)(2k-3)...3$ 
terms.
Wick's theorem is related to the fact that the commutator of the quantum noise
is a c-number; see Footnote \ref{es1}.}.
Due to the factorized structure (\ref{fedor}) of the initial state,
the common averages of $\e $ and various spin operators can be taken
independently.  For example, averaging Eq.~(\ref{k4}) and using Wick's
theorem together with the arithmetic relation $k!\,2^k\,(2k-1)!!=(2k)!$ one
gets:
\BEA
\label{k7}
\langle \sipm (t)\rangle =e^{\pm i\Omega t-iF(t)}
\,\left\langle
\,\pipm(0,t)
\,\right\rangle\,\langle\sipm (0)\rangle
=e^{\pm i\Omega t-\xi(t)}\,\langle\sipm (0)\rangle,
\EEA
where for $t_2\geq t_1$:

\begin{equation}
\begin{split}
\left\langle
\,\pipm(t_1,t_2)
\,\right\rangle
&=\sum _{k=0}^{\infty}\,
\frac{(-1)^{k}}{(2k)!}
\int _{t_1}^{t_2}...\int_{t_1}^{t_2}\, \d s_1...\d s_{2k}\left \langle
{\cal T}\, [\,
\e (s_1)...\e (s_{2k})
\,]
\right \rangle \\
&=\exp \left [
-\frac{1}{2}\,\int_{t_1}^{t_2}\int_{t_1}^{t_2}\,\d s_1\,\d s_2\,
K_{\cal T}(s_1-s_2)
\right ]=\exp[-\xi(t_2-t_1)+i\,F(t_2-t_1)],
\end{split}
\label{ugar} 
\end{equation}
and where

\BEA
\label{xi}
\xi(t)=
\frac{1}{2}\,\int_0^t\int_0^t\,\d s_1\,\d s_2\,
K(s_1-s_2)=
\int_0^t\d s_1\,\int_0^{s_1}\d s_2\,
K(s_2).
\EEA

As seen from (\ref{k7}), $\xi(t)$ characterizes the decay of
$\langle\sipm\rangle$ due to the interaction with the bath. 

\subsection{Ohmic spectrum of the bath.} \label{sec:ohm}

The coupling with the bath can be parametrized via the spectral density
function $J(\om)$:
\BEA
J(\om)=\sum_kg^2_k\, \delta(\om-\om_k).
\label{ja1}
\EEA

In the thermodynamical limit the number of bath oscillators goes to
infinity, and $J(\om)$ becomes a smooth function, whose form is
determined by the underlying physics of the system-bath interaction.

We shall be mainly working with the ohmic spectrum:
\BEA
\label{ja2}
J(\om)=\gamma\,\om\,e^{-\omega/\Gamma}.
\EEA
where $\gamma $ is the dimensionless coupling constant, and where
$\Gamma $ is the maximal characteristic frequency of the bath's
response.  This spectrum and its relevance for describing quantum open
systems was numerously discussed in literature; see, e.g. \cite{rmp}.


\subsubsection{Quantum noise correlation function 
and decay times.}

The correlation function of the quantum noise in the ohmic case, 
using Eqs.(\ref{03}, \ref{ja1}, \ref{ja2}), is given by :
\BEA
K(t)&&=\int _0^{\infty}\d \om\,J(\om)\,\coth
\left[\frac{\hbar\om}{2T}\right]
\cos\omega t\\
&&=\gamma \int _0^{\infty}\d \om\,\om\,\coth
\left[\frac{\hbar\om}{2T}\right]\,e^{-\om /\Gamma}\cos\omega t.
\EEA
Recall that 
the decay factor $\xi (t)$ is related to $K(t)$ via Eq.~(\ref{xi}):
$\ddot{\xi}(t)=K(t)$. Properties of these functions are worked out
in Appendix \ref{ohmo}. In particular, for $\xi(t)$ one gets from
Eqs.~(\ref{a20}, \ref{a30}) the following exact expression:
\BEA
\label{nepal}
\xi (t)
= \gamma \,
\ln \left [\frac{{\bf \Gamma} ^2\left (1+\frac{T}{\hbar\Gamma }
\right )
\,\,\sqrt{1+\Gamma^2t^2}}
{{\bf \Gamma}\left (1+\frac{T}{\hbar\Gamma }
-i\frac{Tt}{\hbar}\right )
{\bf \Gamma} \left (1+\frac{T}{\hbar\Gamma }
+i\frac{Tt}{\hbar}\right )}
\right],
\EEA
where ${\bf \Gamma}$ is Euler's gamma function.
It is seen that the temperature is
controlled by the dimensionless parameter $T/(\hbar\Gamma)$.

Let us now determine the behavior of this quantity for low and large
temperatures.  Using Eq.~(\ref{eon}) one obtains for
$\hbar\Gamma/T\gg 1$ (low temperatures):
\BEA
\label{elleon}
\xi(t)
=\gamma\,\ln \left [
\frac{\hbar\beta}{\pi\,t}\,\sinh \left(\frac{\pi t}{\hbar\beta }\right)
\right]+
\frac{\gamma}{2}\,\ln \left [1+\Gamma^2t^2\right].
\EEA
This implies two regimes of decay: power-law and exponential.
\BEA 
t\ll \hbar\beta :&&
\quad
e^{-\xi(t)}= (1+\Gamma^2 t^2)^{-\gamma/2},\\
t\gtrsim\hbar\beta :&&
\quad
e^{-\xi(t)}= e^{-t/{\cal T}_2}, 
\quad {\cal T}_2=\frac{\hbar}{\gamma T\pi}.
\label{tav1}
\EEA

For $\hbar\Gamma/T\ll 1$ (high temperatures) 
one uses Eq.(\ref{punic}) to get
\BEA
\xi(t)=\frac{2\gamma T}{\hbar\Gamma}\,\left[
\Gamma t\arctan (\Gamma t)-\frac{1}{2}\ln (1+\Gamma^2t^2)
\right].
\EEA
This time the possible regimes of decay can be approximated as
gaussian and exponential.
\BEA 
t\lesssim 1/\Gamma:&&
\quad
e^{-\xi(t)}\simeq e^{-t^2/{\cal T}_2^2}, \quad {\cal T}_2
=\sqrt{\frac{\hbar^2}{\gamma T\Gamma}}, \\
t\gg 1/\Gamma:&& 
\quad 
e^{-\xi(t)}= e^{-t/{\cal T}_2},
\quad {\cal T}_2=\frac{\hbar}{2\gamma T}
\EEA 
In this latter case, as seen from Eq.~(\ref{punic}), $K(t)$ behaves as
approximate delta-function: $K(t)\simeq \frac{ 2\gamma T\Gamma
}{\hbar(1+t^2\Gamma ^2)}$ with the strength $2\gamma T/\hbar $
determined by parameters $\gamma$ and $T$.  Note that in all the above
cases the characteristic times of decay become shorter upon increasing
the temperature $T$ or coupling constant $\gamma$, as is expected. The
gaussian regime of decay was also numeruosly observed in NMR experiments (see
\cite{gauss} and refs. therein). This regime is the basis of the quantum
Zeno effect \cite{petr} and was recently predicted to govern the
reduction process in quantum measurements \cite{cw}. 

\subsubsection{The $G$-factor.}

Finally we will indicate the form of the backreaction functions $G(t)$
and $F(t)$ in the ohmic case (see Eq.(\ref{transilvan})).  As will be seen
below, these functions are rather important for our purposes.
\BEA
\label{castro1}
G(t)&=&\gamma\Gamma
\left(1-\frac{1}{1+\Gamma^2t^2}\right),\\
\label{castro111}
F(t)&=&\gamma
\left[\Gamma t-\arctan(\Gamma t)\right].
\EEA

Since $G(t)$ becomes equal to a constant on the characteristic
time $1/\Gamma$, it is justified to call the latter the
response time of the bath. 

\subsection{Correlated initial conditions.}
\label{corini}

Most papers on the system bath models assume  factorized 
initial conditions. However, in many situations the use
of such a condition is difficult to justify {\it a priori},
since it implies a possibility of switching the system-bath interaction.
Non-factorized initial conditions can be Gibbsians that are 
modified at the initial time, as considered in \cite{NAlinw} for the
Caldeira-Leggett model and in \cite{ANjpa} for the spin-boson model.

For our present purposes it is sensible to use the 
following correlated initial
conditions for the spin and the bath:
\BEA
\label{ka}
\rho(0)=\frac{1}{Z}\exp\left[
-\beta_{\rm S} \HS-\beta (\HI+\HB)
\right],\qquad Z=\tr \,e^{-\beta_{\rm S} \HS-\beta (\HI+\HB)},
\EEA
where $\beta$ 
is the inverse temperature of the bath and 
$\beta_{\rm S}$ is that of the spin. 

The initial condition (\ref{ka}) 
with $\beta_S\not =\beta$ can be generated from
the equilibrium equal-temperature state of the 
overall system via cooling or heating the bath by means of some
superbath. During this process $\siz$ is conserved, and the bath
relaxes to its new temperature under an ``external field'' $\pm \half \X$
generated by the interaction Hamiltonian $\HI$ with $\siz=\pm 1$. More
details of this procedure are given in Appendix \ref{acooling}.

In the thermodynamical limit for the bath, the correlated initial
condition (\ref{ka}) is equivalent to the factorized condition
(\ref{fedor}) with
\BEA
\label{fedor1}
\rho_{\rm S}(0)=\frac{1}{\tr \,e^{-\beta_{\rm S} \HS }}\,
e^{-\beta_{\rm S} \HS},
\EEA
that is, when starting from the factorized initial condition
(\ref{fedor}, \ref{fedor1}), the dynamics of the overall system builds up a
correlated state which at times $t$ much longer than the response time
of the bath, $t\gg 1/\Gamma$~
\footnote{In a more general
model, the one considering $\mathcal{T}_1$, this limit is in fact
$t\gg1/\Gamma$ and $t \ll \mathcal{T}_1$.}
(ergodic limit), is equivalent to (\ref{ka}).  By saying
``equivalent'' we mean that the initial conditions (\ref{fedor}, \ref{fedor1})
and (\ref{ka}) produce the same values for spin's observables and for
collective observables of the bath (i.e., the ones involving summation
over all bath oscillators). This equivalence is further discussed in Appendix 
\ref{maugli}.

As for the initial state of the spin, it can be deduced from Eq.~(\ref{ka})
or from Eq.~(\ref{fedor1})
\begin{equation}
\label{tanh}
\langle\siz\rangle=-\tanh\left[
\frac{\beta_{\rm S}\eps}{2}
\right],\quad 
\langle\six\rangle=
\langle\siy\rangle=0.
\end{equation}

In the following we will use the factorized initial condition
(\ref{fedor}) since it is technically simpler.  The time limit
$t\to\infty$ will be taken before any perturbation acts on the system to
ensure the equivalence with the correlated initial condition
(\ref{ka}).

\renewcommand{\thesection}{\arabic{section}}
\section{Pulsed Dynamics.}
\setcounter{equation}{0}\setcounter{figure}{0} 
\renewcommand{\thesection}{\arabic{section}.}
\label{pulseddynamics}

The external fields acting on the spin are described by 
a time-dependent Hamiltonian
\BEA
\label{hf}
\HF(t)=\half\sum_{k=x,y,z}h_k(t)\,\si_k,
\EEA
with magnitudes $h_k(t)$,
which is to be added to $\HH$ defined in (\ref{ham}) such that the
overall Hamiltonian is time-dependent:
\BEA
\label{hamham}
\HH(t)=\HH+\HF(t).
\EEA 

Eq.~(\ref{hf}) represents the most general external field acting on
the spin. We shall concentrate on {\it pulsed} regime of external
fields which is well known in NMR and ESR physics
\cite{lasers,slichter,nmr,abo,Schmidt,Hahn,Waugh}.  For example, it
was used to describe spin-echo phenomena \cite{Hahn,Waugh} or
processes that switch off undesired interactions, such as those causing
decoherence \cite{slichter,nmr,viola,lidar}.

A pulse of duration $\delta$ is defined by sudden switching on the
external fields at some time $t>0$, and then suddenly switching them off at time
$t+\delta$. It is well-known that during a sudden switching the
density matrix does not change \cite{landau}, while the Hamiltonian 
gets a finite change. Let us for the moment
keep arbitrary the concrete form of external fields in the interval
$(t,t+\delta)$.  The Schr\"odinger evolution operator of the spin+bath from
time zero till some time $t+\tau$, $\tau>\delta$, reads:
\BEA
\label{kapstad}
{\cal T}\exp\left[-\frac{i}{\hbar}\int_0^{t+\tau}\d s\, \HH(s)\right]
&=&e^{-i(t+\tau-t-\delta)\HH/\hbar}
{\cal T}\exp\left[-\frac{i}{\hbar}\int_t^{t+\delta}\d s\, \HH(s)\right]\
e^{-it\HH/\hbar}\\
&=&e^{-i\tau\HH/\hbar}
~\hat{U}_{\rm P}(t)~
e^{-it\HH/\hbar}.
\label{sor}
\EEA
The LHS of Eq.~(\ref{kapstad}) contains the full time-dependent 
Schr\"odinger-representation
Hamiltonian $\HH(s)$, while in the RHS of this equation
we took into account that the actual time-dependence
is present only between $t$ and $t+\delta$. 
The terms $e^{-it\HH/\hbar}$ and $e^{-i(t+\tau-t-\delta)\HH/\hbar}$
stand for the free (unpulsed) evolution in time-intervals
$(0,t)$ and $(t+\delta,t+\tau)$, respectively.
In Eq.~(\ref{sor}) we denoted
\BEA
\label{bom1}
\hat{U}_{\rm P}(t)
&\equiv& e^{i\delta\HH/\hbar}\,
{\cal T}\exp\left[-\frac{i}{\hbar}\int_t^{t+\delta}\d s\,
\HH(s)\right]\\
\label{bom2}
&=&
{\cal T}\exp\left[-\frac{i}{\hbar}\int_t^{t+\delta}\d s\, 
e^{i(s-t)\HH/\hbar}\,\HF(s)
\,e^{i(t-s)\HH/\hbar}
\right]
={\cal T}\exp\left[-\frac{i}{\hbar}\int_0^{\delta}\d s\, 
e^{is\HH/\hbar}\,\HF(s+t)\,e^{-is\HH/\hbar}
\right],
\EEA 
for the pulse evolution operator. The transition from (\ref{bom1})
to (\ref{bom2}) can be made by recalling that $\HH(t)=\HH+\HH_{\rm F}(t)$
and then by noting that the expressions in these
equations satisfy the same first-order
differential equation in $\delta$ with the same boundary condition at 
$\delta=0$ ~\footnote{ In full detail:
\begin{equation}
\begin{split}
\frac{\partial}{\partial \delta}\,\,
e^{i\delta\HH/\hbar}\,
{\cal T}\exp\left[-\frac{i}{\hbar}\int_t^{t+\delta}\d s\,
\HH(s)\right]&=
\frac{i}{\hbar}\left(
\HH-e^{i\delta\HH/\hbar}\,\HH(t+\delta)\,e^{-i\delta\HH/\hbar}\,
\right)\,
e^{i\delta\HH/\hbar}\,
{\cal T}\exp\left[-\frac{i}{\hbar}\int_t^{t+\delta}\d s\,
\HH(s)\right]\nonumber\\ 
&=-\frac{i}{\hbar}
e^{i\delta\HH/\hbar}\,\HH_{\rm F}(t+\delta)\,e^{-i\delta\HH/\hbar}\,
e^{i\delta\HH/\hbar}\,
{\cal T}\exp\left[-\frac{i}{\hbar}\int_t^{t+\delta}\d s\,
\HH(s)\right],
\end{split}
\end{equation}
\begin{equation}
\begin{split}
\frac{\partial}{\partial \delta}\,\,
{\cal T}\exp\left[-\frac{i}{\hbar}\int_0^{\delta}\d s\, 
e^{is\HH/\hbar}\,\HF(s+t)\,e^{-is\HH/\hbar}
\right]&= \nonumber\\
-\frac{i}{\hbar}
e^{i\delta\HH/\hbar}\,\HH_{\rm F}(t+\delta)\,e^{-i\delta\HH/\hbar}
&{\cal T}\exp\left[-\frac{i}{\hbar}\int_0^{\delta}\d s\, 
e^{is\HH/\hbar}\,\HF(s+t)\,e^{-is\HH/\hbar}
\right].
\end{split}
\end{equation}}.

We focus on pulses so short that the influence of the spin Hamiltonian
$\hbar\Omega\siz/2$ and the interaction Hamiltonian $\HI$ can be
neglected during the interval $\delta$. This means that one can take the first
term in the Taylor-expansion ($0<s<\delta$):
\BEA
e^{is\HH/\hbar}\,\HF(s+t)
\,e^{-is\HH/\hbar}&=&\HF(s+t)+\frac{is}{\hbar}\,[\HH,\HF(s+t)]+...\nonumber\\
&=&
\HF(s+t)+\frac{is}{\hbar}\,\left[\frac{\hbar\Omega}{2}\,\siz+\HI,\HF(s+t)\right]+...
\nonumber\\
&\approx& \HF(s+t),
\EEA
Thus, for the pulse evolution operator one gets
\BEA
\hat{U}_{\rm P}(t)=
{\cal T}\exp\left[-\frac{i}{\hbar}\int_0^{\delta}\d s\, 
\HF(s+t)
\right].
\label{papakan}
\EEA
The generalization of the evolution operator
(\ref{sor}) to an arbitrary number of short pulses is straightforward.

Note that in obtaining (\ref{papakan}) we do not require that the bath
Hamiltonian $\HB$ during the pulse is neglected. Since the external
fields are acting on the spin only, the influence of the bath
Hamiltonian disappears by itself from $e^{is\HH/\hbar}\,\HF(s+t)
\,e^{-is\HH/\hbar}$, and is perfectly kept in the general evolution
operator (\ref{kapstad}, \ref{sor}), once the interaction Hamiltonian
$\HI$ has been neglected.

Recalling the orders of magnitude $\hbar\Omega$ and $\hbar\gamma\Gamma$ of
the spin energy and the interaction energy, respectively, |in
particular, recall Eq.~(\ref{kant}), 
$\hbar G=\hbar\int_0^\infty\d \om\,J(\om)/\om$ and
Eq.~(\ref{ja2}),| one gets the following qualitative criteria for the
validity of the short pulsing regime
\BEA
\delta\ll {\rm min}\left(
\Omega^{-1},\, [\gamma\Gamma]^{-1}
\right).
\label{tajik}
\EEA
As it should be, for very small $\gamma$ and a fixed $\Gamma$, the
second restriction on $\delta$ is weaker than the first one.  More
quantitative conditions for the validity of the pulsed regime were
studied recenlty in the context of decoherence suppression by external
pulses \cite{lidar}. 

To deal with the pulsed dynamics in the Heisenberg representation,
one introduces the following superoperators:
\BEA
&&\CE_t\,\hat{A}\equiv e^{i\HH t/\hbar}\,
\hat{A}\,e^{-i\HH t/\hbar},
\\
&&\CP_t\,\hat{A}\equiv 
\hat{U}^\dagger_{\rm P}(t)
\,\hat{A}\,\hat{U}_{\rm P}(t)
\label{hhf}
\EEA
Then the Heisenberg evolution of an operator $\hat{A}$ corresponding to 
Eqs.~(\ref{sor}, \ref{papakan}) reads
\BEA
\label{heisenbergpulse}
\hat{A}(t+\tau)=\CE_t\,\CP_t\,\CE_\tau\,\hat{A}=
e^{i\HH t/\hbar}\,\hat{U}^\dagger_{\rm P}(t)\,
 e^{i\HH \tau/\hbar}\,
   A\,e^{-i\HH \tau/\hbar}\,
     \hat{U}_{\rm P}(t)\,e^{-i\HH t/\hbar}.
\EEA

\subsection{Definition of work.}

The action of external fields on the system is connected with flow of
work. The work done in the time-interval $(0,t)$
is standardly defined as the increase of the average overall
energy of the spin and bath defined by the time-dependent Hamiltonian
$\HH(t)$ \cite{landau,klim,balian}:
\BEA
\label{wo}
W(0,t)=\tr [\,\rho(t)\HH(t)\,]-\tr [\,\rho(0)\HH(0)\,].
\EEA
Due to the conservation of energy of the entire system
(spin+bath+work-source), work is equal to the energy given by the
corresponding work-source (source of external fields).

Since the external fields are acting only the spin, there is a
differential formula for the work which uses only quantities refering
to the local state of the spin and which thus illustrates that the
work-sources exchange energy only through the spin:
\BEA
\label{wowowo}
\frac{\d W}{\d t}=\tr\left(
\rho_{\rm S}(t)\frac{\partial \HH_{\rm F}(t)}{\partial t}
\right),
\EEA
where $\HH_{\rm F}(t)$ as defined by (\ref{hf}) is the contribution of
the external fields into the spin's Hamiltonian, and where $\rho_{\rm
S}(t)$ is the density matrix of the spin.  Eqs.~(\ref{wo},
\ref{wowowo}) relate with each other by the von Neumann equations of
motion 
\BEA
\label{vN}
\dot{\rho}=\frac{i}{\hbar}[\HH(t),\rho(t)]
=\frac{i}{\hbar}[\HH+\HH_{\rm F}(t),\rho(t)] 
\EEA
for the common density
matrix $\rho(t)$ of the spin and the bath, where $\HH$ is the 
Hamiltonian without external fields
\footnote{
In order to get (\ref{wo}) from (\ref{wowowo}), note that the external
fields are acting only on the spin and  $\partial_t\HH_{\rm
F}(t)=\partial_t\HH(t)$. 
Then in expression Eq.~(\ref{wowowo}), we can change the reduced 
density matrix $\rho_{\rm S}$ for the full density matrix $\rho$ 
since the only time dependence of the Hamiltonian lives in the 
Hilbert space of the spin.
Then Eq.~(\ref{wowowo}) can be written as
\BEA
\frac{\d W}{\d t}=\tr\left(
\rho(t)\frac{\partial \HH_{\rm F}(t)}{\partial t}
\right)=
\tr\left(
\rho(t)\frac{\partial \HH(t)}{\partial t}
\right).\nonumber
\EEA
Now integrate this expression from $0$ to $\tau$:
\BEA
\int_0^\tau\d t\,\frac{\d W}{\d t}=
W(0,\tau)=
\int_0^\tau\d t \,\tr\left(
\rho(t)\frac{\partial \HH(t)}{\partial t}\right)=
\tr\left(\rho(\tau)\HH(\tau)\right)-
\tr\left(\rho(0)\HH(0)\right)
-\int_0^\tau\d t \,\tr\left(
\dot{\rho}(t)\,\HH(t)\right).\nonumber
\EEA
Note that the last integral is equal to zero  
due to the equation of motion (\ref{vN}). }.

More specifically, we are interested in the work due to a pulse. 
For the above example of a single pulse at time $t$
this quantity reads from (\ref{sor}, \ref{heisenbergpulse}, \ref{wo}):
\BEA
\label{uu1}
W(0,t+\taut)=W(t,t+\taut)=\tr [\,(\rho(t+\delta)-\rho(t))\,\HH]
=\tr(\rho(t)\,[\CP_t\,\HH-\HH]\,  ).
\EEA

This expression is directly generalized to several successive pulses:
assume that the pulse $\CP_t$ at time $t$ was followed by another
pulse $\CP_{t+\tau}$ at time $t+\tau$ with $\tau>0$. The work done
during the first pulse is given by (\ref{uu1}), while the work done
during the second pulse reads:
\BEA
W(t+\tau,t+\tau+\taut)
=\tr [\,(\rho(t+\tau+\delta)-\rho(t+\tau))\,\HH]
=\tr (\rho(t+\tau) [\CP_{t+\tau}\HH-\HH])\nonumber\\
=\tr(\rho(0)\,\CE_t\CP_t\CE_\tau\,[\CP_{t+\tau}\HH-\HH]).
\EEA

Summing this up with $W(0,t+\taut)$ one gets for the complete work for the 2 pulse situation:
\BEA
W(0,t+\tau+\taut)
=\tr(\rho(0)\,[\CE_t\CP_t\CE_\tau\CP_{t+\tau}\HH-\HH]),
\EEA
as should be.

\subsection{Parametrization of pulses.}

As seen from (\ref{hf}, \ref{hhf}), and taking into account the condition
(\ref{tajik}) which, to all effects, can be taken as $\delta\to 0$, any
pulse corresponds to the most general unitary operation in the Hilbert
space of the spin (this would correspond to a rotation in the classical
language). It is convenient to parametrize pulses 
by coefficients $c_{a,b}$ as:
\BEA
\label{parametrization}
\CP\,\hat{\sigma}_a\equiv 
\hat{U}^\dagger_{\rm P}(t)\,\hat{\sigma}_a
\,\hat{U}_{\rm P}(t)
=\sum_{b=\pm,\,z}c_{a,b}\,\hat{\sigma}_b,
\qquad 
a=\pm,\,z.
\EEA

For more detailed applications we will need the explicit form of
$\hat{U}^\dagger_{\rm P}(t)$ (see (\ref{papakan}))
as a $2\times 2$ unitary matrix whose determinant
can be taken to be unity without loss of generality:
\BEA
\label{para}
\hat{U}^\dagger_{\rm P}(t)
=\left(\begin{array}{rr}
e^{-i\phi}\cos\vartheta & -e^{-i\psi}\sin\vartheta \\
e^{i\psi}\sin\vartheta& e^{i\phi}\cos\vartheta
\end{array}\right),
\EEA
where 
\BEA
0\leq \phi,\,\psi\leq 2\pi,
\qquad
0\leq\vartheta\leq \frac{\pi}{2}.
\EEA

Parametrizations similar to (\ref{para}) are frequently
applied in NMR and ESR experiments
\cite{lasers,slichter,nmr,abo,Schmidt,Hahn,Waugh} where the spin is
rotated certain degrees over a well-defined axis by tuning the parameters
of the laser (microwave) pulse applied.

\renewcommand{\thesection}{\arabic{section}}
\section{Realizations of the model.}
\setcounter{equation}{0}\setcounter{figure}{0} 
\renewcommand{\thesection}{\arabic{section}.}
\label{real}

Once the model with all its ingredients has been defined, we discuss
some of its realizations and provide some numbers. A two-level system
coupled to a thermal bath is a standard model for practically all fields
where quantum systems are studied: NMR, ESR, quantum optics,
spintronics, Josephson junctions, etc. Two particular conditions are,
however, necessary to apply the particular model we study: the condition
${\cal T}_1\gg {\cal T}_2$ on the characteristic relaxation times and
the availability of sufficiently strong pulses. On the other hand, we
can allow for rather short times ${\cal T}^*_2$, since as we will see
this timescale can be overcome with the spin-echo technique. 

There are experimentally realized examples of two-level systems which
have sufficiently long ${\cal T}_2$ times, satisfy in ${\cal T}_1\gg
{\cal T}_2$, e.g. ${\cal T}_1$ exceeds ${\cal T}_2$ by several orders
of magnitude, and admit strong pulses of external fields. For
atoms in optical traps, where ${\cal T}_2\sim 1$s, $1/\Gamma\sim 10^{-8}$s,
there are efficient methods for creating non-equilibrium initial
states and for manipulating atoms by external laser pulses
\cite{atoms}. For an electronic spin injected or optically
excited in a semiconductor, ${\cal T}_2\sim 1\,\mu$s
\cite{spintronics}, and for an exciton created in a quantum
dot ${\cal T}_2\sim 10^{-9}$s \cite{exciton}; in both situations 
$1/\Gamma\sim 10^{-9} - 10^{-13}$s, and femtosecond
($10^{-15}$s) laser pulses are available. In the case of NMR physics
${\cal T}_2\sim 10^{-6}-10^3$s, $1/\Gamma\sim 1 \,\mu$s, and the
duration of pulses can vary between $1 $ps and $1 \,\mu$s 
\cite{slichter,nmr,swedishpnas}.

In all above examples the response time $1/\Gamma$ of the bath is much
shorter than the internal time $1/\Omega$ of the spin. Sometimes it is
argued that such a separation is related to the large size of the bath
and is something generic by itself. This is clearly incorrect, since as
seen from the derivation in section \ref{themodel}, the dimensionless
parameter $\Omega/\Gamma$ has to do with the form of the bath-spin
interaction, rather than with the size of the bath. Moreover, several
examples of bath-spin interaction are known and were analyzed both
experimentally and theoretically, where $\Omega/\Gamma\sim 1$. For
example, Ref.~\cite{nmr1} focusses on relaxation of nuclear spins with
hyperfine frequencies $\Omega \simeq 700$ MHz, ${\cal T}_2< 90$ MHz, and
the ratio $\Omega/\Gamma$ may vary between $10$ and $0.1$. 

Another important parameter that characterizes our setup is the
initial polarization $|\langle\siz\rangle|$ of the spin. It is known
in NMR and ESR physics that the response of magnetic atoms (nucleus)
to external dc magnetic field is best characterized by the $\frac{{\rm
frequency}}{{\rm field}}$ ratio \cite{lasers}, which is for example 
equal to $42$
MHz/T for a proton. For an electron this ratio is $10^3$ times larger
due to the
difference between atomic and nuclear Bohr magnetons,
and for $^{15}$N it is $10$ times smaller. Thus
at temperature $T=1$K and magnetic field $B=1$T the 
equilibrium polarization of a proton is only
$|\langle\si_z\rangle|
=\tanh\frac{\hbar\mu B}{2k_{\rm B}T} = 10^{-3}$, while for an electron it
is $\sim 1$.

\subsection{Exact solution versus various approximations.}

The model as stated above | that is, with the Hamiltonian (\ref{ham},
\ref{hamham}) | is exactly solvable for all temperatures and all
bath-spin coupling constants. It is useful at this point to recall the
reader what are the specific reasons to insist on this feature. The
model with Hamiltonian (\ref{ham}) is a particular case of a more
general spin-boson model, where the influence of ${\cal T}_1$-time is
retained either via an additional term $\propto \six$ in the
Hamiltonian of the spin, or via an additional coupling in the
interaction Hamiltonian. This model is in general not solvable, and
what is worse there are no realiable approximate methods which apply
for a fixed (maybe weak) coupling to the bath and for all temperatures
including the very low ones. The standard weak coupling theories |both
markovian leading to well-known Bloch equations, and non-markovian
ones| are satisfactory only for sufficiently high temperatures, while
at low-temperatures weak-coupling series are singular, and different
methods of their resummation produce different results. In this
context, compare, e.g., convolutionless master equations extensively
discussed in \cite{petr} with a convolutional one worked out in
\cite{div}.

This situation becomes even more problematic under driving by external
fields. The objects studied by us |such as work, energy of the spin|
can be rather fragile to various not very well-controlled
approximations, since there are general limitations governing their
behavior: Thomson's formulation of the second law and restrictions on
work extraction from a two-temperature system (discussed below).
These limitations are derived from the first principles of quantum
mechanics
\cite{thirring,bk,bassett,woron,lenard,ANthomson} and have to be
respected in any particular model.

\renewcommand{\thesection}{\arabic{section}}
\section{General restrictions on work extraction.}
\setcounter{equation}{0}\setcounter{figure}{0} 
\renewcommand{\thesection}{\arabic{section}.}
\label{work-extraction}

The setup of two systems having initially different temperatures and
interacting with a source of work allows to draw a number of general
relations on work-extraction.  Starting from the following general
assumptions:
\begin{enumerate}

\item \emph{Out of equilibrium initial conditions.} 
The initial conditions at the moment $t=0$ are given by Eq.~(\ref{ka}),
where the bath and the spin have initially different temperatures $T$
and $T_{\rm S}$, respectively.  Recall from discussion in section
\ref{corini} that after a small lapse
this initial condition is equivalent to the 
factorized one (\ref{fedor}, \ref{fedor1}).
We use the former one since it is more convenient when dealing with 
the general restrictions on the work-extraction.

\item \emph{Cyclic external fields.}
For the following derivation, the Hamiltonian $\HH_{\rm F}(t)$ 
of external fields acting on 
the spin is completely arbitrary. In particular, it need not be
composed by pulses, where it would vanish outside of the
pulses. The only general assumption made on 
$\HH_{\rm F}(t)$ is that its action is cyclic at some final time
$t_{\rm f}$:
\BEA 
\label{cy}
\HH_{\rm F}(0)=
\HH_{\rm F}(t_{\rm f})=0.
\EEA

\end{enumerate}

We can find the following two relations (derived explicitly in 
Appendix \ref{acarnot}):
\BEA
\label{fa1}
&&W\geq \left(1-\frac{T}{T_S}\right)\,\Delta H_{\rm S},\\
\label{fa2}
&&W\geq \left(1-\frac{T_S}{T}\right)\,(\Delta H_{\rm I}+\Delta H_{\rm B}),
\EEA
where 
\BEA
\Delta\HH_{\rm k}=\tr\left (\HH_{k}\, [\,
\rho(t_{\rm f})-\rho(0)\,]
\right), \qquad {\rm k}={\rm S}, {\rm I}, {\rm B},
\EEA
are the changes of the corresponding average energies of the spin, bath
and interaction, with
$\rho(t_{\rm f})$ being the complete density matrix of the spin and bath
at time $\tau$, and where
the total work reads:
\BEA
W=\Delta H_{\rm S}+\Delta H_{\rm I}+\Delta H_{\rm B}.
\label{mimi}
\EEA
Here are implications of Eqs.(\ref{fa1},\ref{fa2}).

\begin{itemize}
\item
If $T_S>T$ and work is extracted, $W<0$, (\ref{fa1}) implies 
\BEA
\label{batrak1}
\Delta H_{\rm S}<0,\qquad
\Delta H_{\rm I}+\Delta H_{\rm B}>0:
\EEA
the system looses energy, while the bath gains it and the amount of 
the extracted work $|W|$ is then bounded from above
by $|\Delta H_{\rm S}|$.

\item
If $T=T_S$, both Eqs.(\ref{fa1}, \ref{fa2}) produce: $W\geq 0$, which
is, in fact, the statement of the second law in Thomson's formulation:
no work can be extracted from an equilibrium system by means of cyclic
perturbations.

\item
If $T_S<T$, inequalities in Eq.(\ref{batrak1}) are reversed: now
work-extraction implies that 
\BEA
\label{batrak2}
\Delta H_{\rm S}>0,\qquad
\Delta H_{\rm I}+\Delta H_{\rm B}<0,
\EEA
and $|W|$ is then bounded from above
by $|\Delta H_{\rm I}+\Delta H_{\rm B}|$.

\end{itemize}

These conclusions are close to what one could have expected
from the standard (phenomenological) thermodynamical reasoning \cite{landau}.  
However, it should be emphasized that in contrast to typical textbook
derivations, Eqs.~(\ref{fa1}, \ref{fa2}) were derived starting from 
first principles (see Appendix \ref{acarnot}), and, moreover, their
derivation is by no means restricted to a weak bath-spin coupling, a
condition which need not be satisfied in practice.

\subsection{Efficiency and Heat.}

Another useful notion is the efficiency $\eta$ of the work-extraction,
which shows how economically 
non-equilibrium, two-temperature resource is employed in 
work-extraction \cite{landau,klim,balian}.
The special importance of efficiency is related to the fact that in the
standard thermodynamics it is bounded from above by Carnot's value,
which is a system-independent quantity.

Though our system starts out of equilibrium due to different {\it
initial} temperatures of the spin and the bath, the notion of
efficiency should be studied for it anew, since it does not
automatically fall into the class of heat-engine models, as studied in
textbooks of thermodynamics and statistical physics
\cite{landau,klim,balian}:
\begin{itemize}

\item
There is no working body which operates cyclically between two thermal
baths. With us cyclic processes are defined with respect to the work
source.

\item The interaction between the systems having different temperatures
|in the case discussed here, the spin and the bath| need not be weak.

\item We do not require that our systems always stay very close to
equilibrium.  In contrast, both during and immediately after the
work-extraction process, the spin is in a non-equilibrium state, which
in general cannot be described in terms of a time-dependent
temperature.

\end{itemize}

However, in spite of all these differences we can define the notion of
effciency and this will be an equally useful characterization of the
work-extraction process \cite{landau,klim,balian}.

Recall that external fields are acting exclusively on the spin
variables and not on those of the bath. This implies that when during
work-extraction the source of work receives energy $|W|$, this energy
consists of a contribution coming directly from the spin and of a part
which comes to the work-source from the bath but through the spin. 
In this context one can write the change of energy of the spin as

\begin{equation}
\begin{split}
\frac{\d}{\d t} \, &\tr\left(\,\rho_{\rm S}(t)\,\HS(t)
\,\right) \\
&=\tr\left[\,\left(\frac{\d}{\d t}\rho_{\rm S}(t)\right)\,\HH_{\rm S}(t)
\,\right]
+ \tr\left[\,\rho_{\rm S}(t)\,\left(\frac{\partial}{\partial t} 
\HH_{\rm S}(t)\right)\,\right] = \frac{\d}{\d t} Q+\frac{\d}{\d t} W,
\label{eq:incre}
\end{split}
\end{equation}
where in our case the Hamiltonian of the spin reads from Eqs. 
(\ref{hamspin}, \ref{hf}) (note analogy with (\ref{hamham})): 
\BEA
\HH_{\rm S}(t)=\frac{\eps}{2}\siz+
\half\sum_{k=x,y,z}h_k(t)\,\si_k.
\EEA 
The partial time-derivative in (\ref{eq:incre}) stresses that we are
in Schr\"odinger representation.  When deriving (\ref{eq:incre}) we
have used $\partial_t \HH_{\rm S}(t)=\partial_t \HH_{\rm F}(t)$ and
(\ref{wowowo}).  The last equality in (\ref{eq:incre}) serves as a
definition of heat ($\d Q$) ~\footnote{Note that in the equilibrium
thermodynamics people frequently distinguish functions of a
quasi-equilibrium thermodynamical process (the one which can be viewed
as a chain of equilibrium states) from functions of the state. In this
context the change in heat is written as $\dbarrm Q$. Here we consider
(possibly strongly) non-equilibrium situations, where almost any
quantity (e.g. energy) is a function of the process. Therefore, we do
not introduce the symbol $\dbarrm$. }.

Integrating this from 0 to $\tau$ and using (\ref{cy}) and
(\ref{mimi}) we obtain
\BEA
\Delta Q=-\left(\Delta H_{\rm I}+
\Delta H_{\rm B}\right).
\EEA

Note that in the above definition of heat, the average interaction
energy is attributed to the heat received from the bath altough it
by itself depends also on the variables of the spin; see Eq.~(\ref{hamint}).
The reason for this asymetry is clearly contained in the very
initial statement of the problem, where we |quite in accordance with
the usual practice of statistical physics| restricted the work source
to act only on the spin.

All this being said, one can now proceed for $W<0$ (work-extraction)
with the usual definition of efficiency as the ratio of the useful
energy $|W|$ to the maximal energy involved in the work-extraction:
\BEA
\label{takanq}
\eta\equiv\frac{|W|}
{{\rm max}\left(\,
|\Delta H_{\rm S}|, \,|\Delta H_{\rm I}+\Delta H_{\rm B}|\,\right)}.
\EEA

For $T_{\rm S}>T$ Eqs.~(\ref{mimi}, \ref{batrak1}) and $W<0$
imply $|W|=|\Delta H_{\rm S}|-|\Delta H_{\rm I}+\Delta H_{\rm B}|$,
and then (\ref{takanq}) results in
\BEA
\label{takanq1}
\eta=\frac{|W|}{|\Delta H_{\rm S}|}.
\EEA

Analogously, for  $T_{\rm S}<T$ we have
\BEA
\label{takanq2}
\eta=\frac{|W|}{|\Delta H_{\rm I}+\Delta H_{\rm B}|}=
\frac{|W|}{|W|+|\Delta H_{\rm S}|},
\EEA
from $|W|=|\Delta H_{\rm I}+\Delta H_{\rm B}|-|\Delta H_{\rm S}|$.

It is now seen from 
Eqs.~(\ref{fa1}, \ref{fa2}, \ref{batrak1}, \ref{batrak2}) 
that the efficiency is always
bounded by the Carnot value:
\BEA
\label{takanq3}
\eta\leq
1-\frac{{\rm min}\left(\,T,\,T_{\rm S}\,\right)}
{{\rm max}\left(\,T,\,T_{\rm S}\,\right)}.
\EEA

\renewcommand{\thesection}{\arabic{section}}
\section{Work-extraction via two pulses.}
\setcounter{equation}{0}\setcounter{figure}{0} 
\renewcommand{\thesection}{\arabic{section}.}
\label{2pulses}

\subsection{Setup of pulsing.}

Let us now detailze the setup of work-extraction. The spin and the
bath are prepared in the state (\ref{ka}) with different temperatures
$T_{\rm S}$ and $T$ for the spin and the bath, respectively.  Thus,
the initial average population difference $\langle\siz\rangle$ is
given by (\ref{tanh}).

Alternatively, we can prepare the spin+bath in the state (\ref{fedor},
\ref{fedor1}).  In this case, one waits for a time $t\gg 1/\Gamma$
for ensuring the robustness of the results. Then the setup
does not depend on details of the initial preparation, because the
initial conditions (\ref{fedor}, \ref{fedor1}) and (\ref{ka}) have become
equivalent. 

The final ingredient of the setup are pulses 
\BEA \CP_{t}=\CP_1, \qquad\CP_{t+\tau}=\CP_2,
\EEA 
applied at times $t$ and $t+\tau$, respectively.

\subsection{Formulas for work.}

The work done for the first pulse reads (as defined in Eqs.(\ref{uu1})):
\BEA
\label{w1}
W_1=\frac{\eps}{2}\left\langle
\CP_1\siz-\siz
\right\rangle_{t}+
\frac{\hbar}{2}\left\langle\left(
\CP_1\siz-\siz\right)\X
\right\rangle_{t},
\EEA
where for any operator $\hat{A}$ the average
\BEA
\left\langle \hat{A}\right\rangle_{t}=\tr[\, \hat{A}\,\rho(t)\,]
\EEA
refers to the time $t$ just before the application of the pulse.
The value of $W_1$ is worked out by recalling the parametrization
(\ref{para}), the evolution of the collective bath coordinate $\X$ as 
given by Eq.(\ref{kant}), and finally the initial condition
(\ref{fedor}, \ref{fedor1}). The final result reads:
\BEA
\label{w11}
W_1=(\,1-c^{(1)}_{z,z}\,)\left [ 
\frac{\hbar}{2}\,G-\frac{\eps}{2}\langle\siz\rangle \right], 
\EEA 
with $G$ as defined in Eq. (\ref{shopen}), and where
$c^{(1)}_{z,z}$ is the corresponding parametrization coefficient of
the first pulse as defined by 
(\ref{parametrization}).

As follows from $T_{\rm S} >0$ and $\langle\siz\rangle<0$
(see (\ref{tanh})), the
work $W_1$ is always positive. This is in agreement with the
thermodynamical wisdom of local equilibrium: the second term in the
RHS of Eq. (\ref{w11}) is the contribution from the spin energy and it
is positive, since the spin was in equilibrium before the application
of the first pulse.  Another positive term $\half
(\,1-c^{(1)}_{z,z}\,)\,\hbar\,G$ in the RHS of Eq. (\ref{w11}) comes
from the interaction Hamiltonian (the bath operators, and thus the
bath Hamiltonian, are not influenced by this first pulse). Again, it
is intuitively expected that the interaction Hamiltonian should make
the average energy costs higher.

The work done for the second pulse reads analogously to Eq. (\ref{w1}):
\BEA
W_2=\frac{\eps}{2}\left\langle
\CP_2\siz-\siz
\right\rangle_{t+\tau}+
\frac{\hbar}{2}\left\langle\left(
\CP_2\siz-\siz\right)\X
\right\rangle_{t+\tau},
\label{2pul}
\EEA
where the averages $\langle ...\rangle_{t+\tau}$ refer to the
time just before the application of the second pulse.

Eq.~(\ref{2pul}) is worked out in Appendix \ref{twopulses}
with the result for the total work $W=W_1+W_2$ being:
\BEA
\label{w2}
W=&&-\frac{\eps}{2}\,\left(1-c^{(2)}_{z,z}\,c^{(1)}_{z,z}\,\right)\,
\langle\siz\rangle
+\eps\,e^{-\xi(\tau)}\,
\Re\left\{c^{(1)}_{+,z}\,c^{(2)}_{z,+}\,
e^{i\Omega \tau}\,\left\langle e^{i\chi_2\,\siz}\,\siz\right\rangle\right\}
\\
&&+\frac{\hbar\, G}{2}\,(1-c^{(1)}_{z,z})
+\frac{\hbar}{2}\,(1-c^{(2)}_{z,z})\left(
G(\tau)+g_2(\tau)\,c^{(1)}_{z,z}\right)
\label{w222}
\\
&&+e^{-\xi(\tau)}\,\Re\left\{c^{(1)}_{+,z}\,
c^{(2)}_{z,+}\,e^{i\Omega\tau}\,
\left(
i\hbar\,\dot{\xi}(\tau)\,\left\langle e^{i\chi_2\,\siz}\siz\right\rangle
+\hbar\,
g_2(\tau)\,\left\langle e^{i\chi_2\,\siz}\right\rangle
\right)\right\}.
\label{w22}
\EEA

The detailed explanation of various terms in this expression
and of their physical meaning comes as follows. 

The first term in the RHS of Eq. (\ref{w2}) is the contribution from the 
initial spin
energy. The second term comes from the transversal degrees of
freedom excited by the first pulse. The factor
$e^{-\xi(\tau)}$ accounts for the reduction of these terms in the time
interval $\tau$ between pulses. Recall that the parametrization coefficients
$c^{(1,2)}_{+,z}$ and $c^{(1,2)}_{z,z}$ for the first and the second 
pulse are defined in Eq.~(\ref{parametrization}):

The terms in Eqs. (\ref{w222}, \ref{w22}) are the joint contribution from
the bath Hamiltonian (\ref{hambath}) and from the interaction
Hamiltonian (\ref{hamint}).  The last of them couples to the
transversal degrees of the spin, as reflected by the presence of
$e^{-\xi(\tau)}$.  Recall that the averages $\langle ...\rangle$ in
Eqs. (\ref{w2}, \ref{w22}) refer to the initial state Eqs. (\ref{fedor},
\ref{fedor1}). Finally, the factors 
\begin{gather}
\label{g2}
g_2(\tau)\equiv G-G(\tau)
=\frac{\gamma\Gamma}{1+\tau^2\Gamma^2},\\
\label{chi2}
\chi_2(\tau)=
-\gamma\arctan(\tau \Gamma),
\end{gather}
(the lower index $2$ refers to the two-pulse situation) come from the
backreaction of the spin to the bath.  

Next we note that the behavior of $W=W_1+W_2$ is controlled by five
dimensionless parameters (see Appendix \ref{twopulses}), which for 
the ohmic case reads
\BEA
W=W_1+W_2=\frac{\hbar\gamma \Gamma}{2}\, w\left(
\frac{T}{\hbar\Gamma},\,\gamma, \,
\frac{\eps}{\hbar\Gamma}, \,\langle \siz\rangle,
\,\tau\Gamma\right).
\label{dimo}
\EEA
Note that the spin temperature
$T_{\rm S}$ enters only through the initial 
(at $t=0$) $\langle \siz\rangle$ 
as given by Eq.~(\ref{tanh}).

There are two situations within the present setup, where
work-extraction is not prohibited: $T>T_{\rm S}$ and $T<T_{\rm S}$. We
deal with them separately, since for these cases the work-extraction
effect exists in different ranges of the parameters.

\subsection{Work extraction for $T>T_{\rm S}$.}

\begin{figure}[ht] 
\includegraphics[width=7cm]{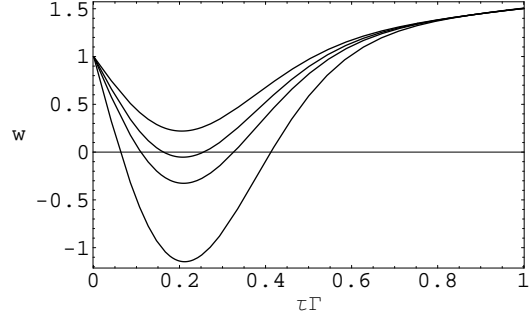} 
\caption{Dimensionless total work $w$ (see Eq. (\ref{dimo}) in the text) 
versus dimensionless time $\tau\Gamma$
(the waiting time between the two pulses) in the regime $T>T_{\rm S}$.
We compare the extracted work for different values of the initial
polarization (or equivalently, of the initial temperature) of the spin.
$\frac{T}{\hbar\Gamma}=10$, $\gamma=1$,
$\frac{\eps}{\hbar\Gamma}=0.01$, $\langle\siz\rangle=-0.8,
\,-0.5,\,-0.4,\,-0.3$ (from bottom to top). The two pulses are given by
Eqs.~(\ref{pulse1}, \ref{pulse2}, \ref{pp}).
Work-extraction disappears for larger 
$\langle\siz\rangle$, that is, for closer (initial) temperatures
of the spin and the bath.
} 
\label{w2p1} 
\end{figure}

\begin{figure}[ht] 
\includegraphics[width=7cm]{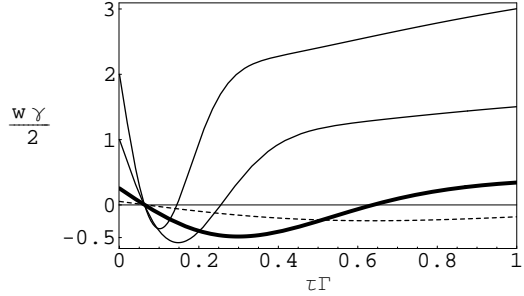} 
\caption{The ratio $\frac{W}{\hbar\Gamma}=\frac{w\gamma}{2}$
(see Eq. (\ref{dimo}) in the text)
versus the dimensionless time $\tau\Gamma$
for two pulses in the regime $T>T_{\rm S}$. 
We compare the extracted work for different values of the
dimensionless bath-spin coupling constant $\gamma$.
$\frac{T}{\hbar\Gamma}=10$, 
$\frac{\eps}{\hbar\Gamma}=0.01$, $\langle\siz\rangle=-0.8$ and
$\gamma=4$ (upper solid curve),
$\gamma=2$ (lower solid curve),
$\gamma=0.5$ (bold curve),
$\gamma=0.1$ (dotted curve),
The two pulses are given by
Eqs.~(\ref{pulse1}, \ref{pulse2}, \ref{pp}). It is seen that
the maximal extracted work is a non-monotonous
function of the dimensionless coupling constant ${\gamma}$.
} 
\label{w2p11} 
\end{figure} 

It was seen above that the first pulse always costs work, since it is
applied on the spin whose state is (initially) in local equilibrium
at temperature $T_{\rm S}$. However, the first pulse can do more than
simply wasting work. Consider, for example, a $\frac{\pi}{2}$ pulse 
in the $y$-direction
\footnote{Here and after, we do not present the results 
of the  full optimization of the work over the parameters of the
involved pulses. The reason is that we
do not want to make the pulsing setup too much dependent on the
details of the model. On the other hand, the results of this full
optimization did not show any qualitative difference with the
presented ones.}:
\BEA
\label{pulse1}
\CP_1=\CP\left(\frac{\pi}{2};y\right),
\EEA
where
\BEA
&&\CP\left(\varphi;y\right)\,\siz\equiv
e^{i\,\varphi\siy/2}\,\si_z\,e^{-i\,\varphi\siy/2}
=\siz\cos \varphi -\six\sin \varphi ,\\
&&\CP\left(\varphi;y\right)\,\six\equiv
e^{i\,\varphi\siy/2}\,\six\,e^{-i\,\varphi\siy/2}=
\siz\sin \varphi +\six\cos \varphi ,\\
&&
\CP\left(\varphi;y\right)\,\siy\equiv\siy .
\label{tarzan1}
\EEA 
This pulses excites the transversal component $\langle\six\rangle$
which starts to decay under action of the bath, and thus correlations
between the spin and the bath are established. The proper second pulse
is then applied at time $\tau$, for instance $+\frac{\pi}{2}$ in the
$x$-direction:
\BEA
\label{pulse2}
\CP_2=\CP\left(\frac{\pi}{2};x\right),
\EEA
where
\BEA
&&\CP\left(\varphi;x\right)\,\siz\equiv
e^{i\,\varphi\six/2}\,\si_z\,e^{-i\,\varphi\six/2}
=\si _z\cos \varphi +\si _y\sin \varphi ,\\
&&\CP\left(\varphi;x\right)\,\siy\equiv
e^{i\,\varphi\si_x/2}\,\siy\,e^{-i\,\varphi\six/2}=
-\siz\sin \varphi +\siy\cos \varphi ,\\
&&\CP\left(\varphi;x\right)\,\six\equiv\six.
\label{tarzan2}
\EEA 

Note that our choice of pulses corresponds to
\BEA
\label{pp}
c^{(1)}_{+,z}=1,\quad
c^{(2)}_{z,+}=\frac{1}{2i}, \quad 
c^{(1)}_{z,z}=0, \quad c^{(2)}_{z,z}=0. 
\EEA

It appears that not only some work is extracted by the second
pulse, but the overall work by the two pulses can be negative
for properly chosen time $\tau$:
\BEA
W=W_1+W_2<0, 
\EEA
as seen in Figs.~\ref{w2p1}, \ref{w2p11}.  
This is one of the central results of this paper.

The time $\tau$ needed for work-extraction should be neither too short
|otherwise the two pulses will effectively sum into one, and we know
that no work-extraction is achieved by a single pulse,| nor too long,
otherwise the transveral degree of freedom excited by the first pulse
will decay, and we will have two isolated single pulses. This is seen
in Figs.~\ref{w2p1}, \ref{w2p11}. Note that the choice of pulses is
obviously important for having work-extraction.  Eqs.~(\ref{pulse1},
\ref{pulse2}) represent only one particular example leading to
work-extraction in the regime $T>T_{\rm S}$.  

As for the magnitude of the extracted work, one notes from
Eq.~(\ref{dimo}) and Fig.~\ref{w2p1} that it is of order of
$\hbar\Gamma/2$, which is basically the response energy of the bath.
This is not occasional, since as seen from (\ref{batrak2}), the work in
this regime $T>T_{\rm S}$ is coming from the bath.

Noting the ratio $\eps/(\hbar\Gamma)=0.01$ in Fig.~\ref{w2p1} |this
and even smaller ratios are usual for the realizations of the model as
we discussed in section \ref{real}| we conclude that the extracted
work can be of several orders of magnitude larger than the energy of
the spin. On the other hand, the extracted work is limited by $\sim T$
which is the characteristic thermal energy available in the bath.
Indeed, as seen in Fig.~\ref{w2p1} the etracted work can be of order
of $\hbar\Gamma$, while the bath temperature is nearly ten times
larger: $T=10\,\hbar\Gamma$. Not unexpectedly, work-extraction
disappears when the temperatures $T$ and $T_{\rm S}$ are close to each
other; see Fig.~\ref{w2p1}.

Let us return once again to the optimal time-interval $\tau$.
As Figs.~\ref{w2p1},
\ref{w2p11} show, the value of $\tau$ 
at which the extracted work is maximal is roughly of the same order of
magnitude as $1/\Gamma$. However, the optimal $\tau$ can be much
larger (e.g., $\sim 10^3/\Gamma$) for smaller coupling constants
$\gamma$, that is, one can increase the waiting time between the
pulses at the expense of reducing the magnitude
$\propto\hbar\gamma\Gamma$ of the extracted work.

\subsection{Work extraction for $T<T_{\rm S}$.}

\begin{figure}[ht] 
\includegraphics[width=7cm]{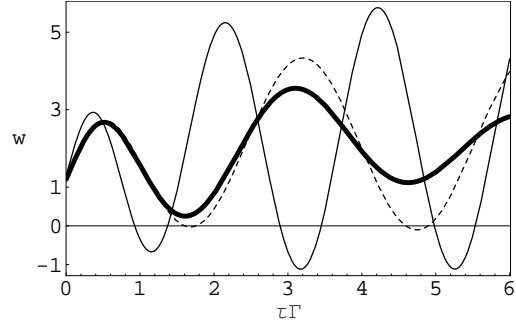} 
\caption{Dimensionless work $w$ (see (\ref{dimo}))
versus dimensionless time $\tau\Gamma$
for two pulses in the regime $T<T_{\rm S}$ in the case of
$\gamma=0.1$, $\langle\siz\rangle=-0.01$. The choice of pulses is
given by Eqs.~(\ref{ppp3}, \ref{ppp4}).
Full line: $\frac{T}{\hbar\Gamma}=0.1$, $\frac{\eps}{\hbar\Gamma}=3$, 
Dashed line: $\frac{T}{\hbar\Gamma}=0.1$, $\frac{\eps}{\hbar\Gamma}=2$.
Bold line: $\frac{T}{\hbar\Gamma}=1$, $\frac{\eps}{\hbar\Gamma}=3$.
} 
\label{w2p2} 
\end{figure}

Let us now turn to scenarios of work-extraction in the regime $T_{\rm
S} >T$. As seen from (\ref{batrak1}),  
if there is work-extraction at all in this regime, the work should
come from the average energy difference of the spin, while 
$\Delta H_{\rm I}+\Delta H_{\rm B}$ is then necessarily
positive.  Since the latter quantity is of order of $\gamma\Gamma$
(response energy of the bath), and the spin's energy difference is
obviously of order $\eps$, there are two ways to try to achieve
work-extraction, that is, to get $W= |\Delta H_{\rm I}+\Delta H_{\rm
B}| -|\Delta H_{\rm S}|<0$: One should either take
$\eps/(\hbar\Gamma)\sim 1$ or take the dimensionless coupling constant
$\gamma$ very small.  The second way did not lead to
work-extraction, since the required coupling constants are so small
that the spin effectively decouples from the bath. In contrast, the
first case with $\eps/(\hbar\Gamma)\sim 1$ led to a sizable
work-extraction, as see in Fig.~\ref{w2p2}. Recall in this context
that systems with $\eps/(\hbar\Gamma)=\Omega/\Gamma\sim 1$ are
well-known; see section \ref{real} for details.

As compared to the previous regime, here the choice of pulses has
to be different for the work-extraction to be possible. For example,
\BEA
\label{ppp3}
\CP_1=\CP\left(-\frac{\pi}{2};x\right),
\qquad
\CP_2=\CP\left(-\frac{\pi}{2};y\right),
\EEA
for the first and the second pulses respectively; see 
Eqs.~(\ref{tarzan1}, \ref{tarzan2})for the definitions of pulses. 
We see from (\ref{para}) that this choice amounts to substituting 
\BEA
\label{ppp4}
c^{(1)}_{+,z}=i, 
\quad
c^{(2)}_{z,+}=\frac{1}{2}, \quad
c^{(1)}_{z,z}=0,\quad 
c^{(2)}_{z,z}=0,
\EEA
into Eqs.~(\ref{w2}, \ref{w222}, \ref{w22}). 

\subsection{Efficiency of work extraction.}

\begin{figure}[ht] 
\includegraphics[width=7cm]{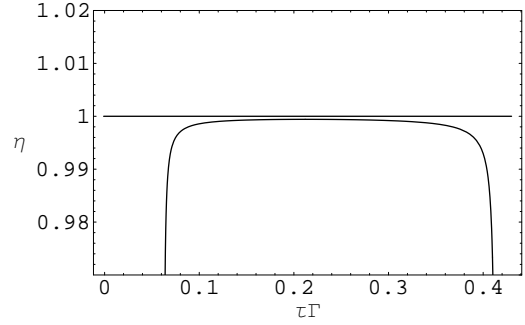} 
\caption{Efficiency $\eta$
versus dimensionless time $\tau\Gamma$
for two pulses in the regime $T>T_{\rm S}$.
$\frac{T}{\hbar\Gamma}=10$, $\gamma=1$,
$\frac{\eps}{\hbar\Gamma}=0.01$, $\langle\siz\rangle=-0.8$.
The two pulses are given by
Eqs.~(\ref{pulse1}, \ref{pulse2}, \ref{pp}).
The efficiency is slightly below than the corresponding Carnot's value 
and is maximized over $\tau\Gamma$ almost simultaneously 
with the dimensionless work $w$; see Fig.~\ref{w2p1}.
} 
\label{ef2p1} 
\end{figure}

\begin{figure}[ht] 
\includegraphics[width=7cm]{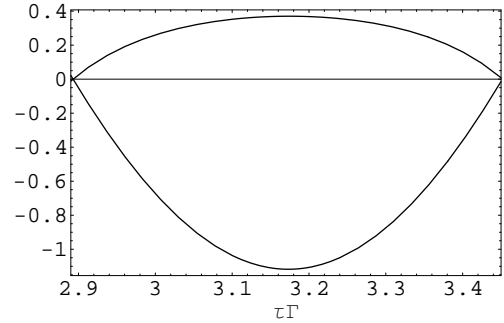} 
\caption{Efficiency $\eta$ (upper curve) and dimensionless work
$w$ (lower curve)
versus dimensionless time $\tau\Gamma$
for two pulses in the regime $T<T_{\rm S}$.
$\frac{T}{\hbar\Gamma}=0.1$, $\gamma=0.1$,
$\frac{\eps}{\hbar\Gamma}=3$, $\langle\siz\rangle=-0.01$.
The two pulses are given by Eqs.~(\ref{ppp3}, \ref{ppp4}).
The efficiency is below than the corresponding Carnot's value $0.99$
and is maximized over $\tau\Gamma$ almost simultaneously 
with the dimensionless work $w$.
} 
\label{ef2p2} 
\end{figure} 

We shall now discuss the efficiency of work-extraction as defined by
Eqs.~(\ref{takanq}, \ref{takanq1}, \ref{takanq2}).  To calculate it
one needs to know the total work given by Eqs.~(\ref{w11}, 
\ref{w2}--\ref{w22}), and the contribution $\Delta H_{\rm S}$ 
to the work $W$ coming from the average energy of the spin, 
which is read from the RHS of Eq.~(\ref{w2}).

The efficiency as a function of $\tau\Gamma$ is presented by
Figs.~\ref{ef2p1}, \ref{ef2p2} for $T>T_{\rm S}$ and $T<T_{\rm S}$,
respectively.  There are several important things to note.  
\begin{itemize}

\item  For
$T>T_{\rm S}$ the efficiency can be very close to unity, if the
temperatures $T$ and $T_{\rm S}$ are sufficiently separated from each
other, which is the case in Fig.~\ref{ef2p1}. It is, however, always
limited by Carnot's value, as given by Eq.~(\ref{takanq3}).  For
$T<T_{\rm S}$ the effiency is sizable, but is rather below the
corresponding Carnot value.  

\item
The work and efficiency are
maximized over $\tau\Gamma$ {\it simultaneously}.

\item
Recall in this context that in the standard
thermodynamics efficiencies close to the optimal value are connected
to very small work per unit of time (zero power of work), since they
are achievable for very slow processes. This is not the case with the
presented setup. As seen from Figs.~\ref{w2p1}, \ref{w2p11},
\ref{w2p2}, the work is extracted on times which are of order of
$1/\Gamma$ (response time of the bath), which is typicaly much smaller
than the internal characteristic time $1/\Omega$ of the spin. Thus, in
Fig.~\ref{ef2p1} we have nearly optimal efficiencies together with the
maximal work and a finite power of work.

\end{itemize}

\renewcommand{\thesection}{\arabic{section}}
\section{Work-extraction via spin-echo pulses.} 
\setcounter{equation}{0}\setcounter{figure}{0} 
\renewcommand{\thesection}{\arabic{section}.}
\label{spinecho}

So far we assumed that we deal either with a single spin coupled to the
bath, or, equivalently, with an ensemble of identical non-interacting
spins each coupled with its own bath \footnote{The assumption that each
spin has its bath is a natural one for cases when the spins are
sufficiently well separated \cite{slichter,nmr}. This assumption is in a
sense also a pessimistic one, since admitting a single bath for all the
involved spins |a situation which has its own relevance in NMR/ESR
physics \cite{abo}| we may get additional, collective channels of
work-extraction.}.  However, many experiments |especially in NMR
physics| are done on ensembles of non-interacting spins which are not
in identical environment. The difference lies in the different 
energies $\eps$. This can
be caused by inhomogeneous fields contributing into energy $\eps$, or
by action of environment, e.g., chemical shifts for nuclear spins
\cite{slichter,nmr,abo} or effective g-factors for electronic spins in
a quantum dot.  
It is customary to regard these energies as random quantities, so that the
collective outcomes from such ensembles are obtained by averaging over
$\eps=\hbar\Omega$ the corresponding expressions for a single spin. 
We shall assume that the distribution of $\Omega$ is gaussian with
average $\Omega_0$ and dispersion $d$:
\footnote{The assumption on the gaussian character of this distribution
can be motivated by the central limit theorem, where the randomness of
$\Omega$ is viewed to be caused by many (nearly) independent 
small random factors.}
\BEA
P(\Omega)=\frac{1}{\sqrt{2\pi d}}\,e^{-(\Omega-\Omega_0)^2/(2d)}.
\label{disorder}
\EEA 

It is now clear that the averaging over $P(\Omega)$ the oscillating terms
$e^{i\Omega\tau}$ will produce $\sim e^{-d\tau^2/2}$ that is, a strong
decay on characteristic times 
\BEA
{\cal T}_2^*\propto 1/\sqrt{d}. 
\EEA

For $\tau/{\cal T}_2^*\gg 1$ all the terms containing $e^{i\Omega\tau}$
will be zero after averaging, and the corresponding averaged work for
two pulses will always be positive as seen from Eqs. (\ref{w11}, \ref{w2}, \ref{w222},
\ref{w22}).  Indeed, all possible negative values of the full work $W$
were related to transversal degrees of freedom excited by the first
pulse. These terms come with the factor $e^{i\Omega\tau}$ which is
connected to the free evolution in the time-interval $\tau$ between
the two pulses. Due to the decay of these terms after $\tau/{\cal
T}_2^*\gg 1$, it is impossible to extract work from this ensemble via
two pulses.

However, we can extract work even in the
strongly-disordered situation with ${\cal T}_2^*$ being short, if we
combine our work-extraction setup with the spin-echo phenomenon
\cite{Hahn,Waugh}.  For our present purposes this amounts to applying
a $\pi$-pulse, for instance in $x$-direction:
\BEA
\label{pi}
\CP_\pi\,\siz=-\siz,\quad 
\CP_\pi\,\siy=-\siy,\quad
\CP_\pi\,\six=\six,
\EEA 
right in the middle of two pulses $\CP_1$ and $\CP_2$ 
(to be tuned later on) applied at times
$t$ and $t+2\tau$, respectively.  The work done by the first pulse
reads from Eq. (\ref{w11}) after averaging over $P(\Omega)$ given by
Eq. (\ref{disorder}):
\BEA
W_1=(\,1-c^{(1)}_{z,z}\,)\left [ \frac{\hbar}{2}
\,G-E
\right], 
\EEA
where
\BEA
E=-
\frac{\hbar}{2}\int\d 
\Omega\, P(\Omega)\Omega\tanh\frac{\beta_{\rm S}\hbar\Omega}{2}<0,
\EEA
is the average initial energy of the ensemble of spins. The work done by the 
$\pi$-pulse at time $t+\tau$
is found from (\ref{2pul}) by substituting there the
parameters $c^{(2)}_{z,z}=-1$ and $c^{(2)}_{z,+}=0$ of this pulse:
\BEA
W_\pi=\hbar\,G(\tau)+\hbar\,g_2(\tau)\,c^{(1)}_{z,z}
-2E\,c^{(1)}_{z,z},
\EEA
where $g_2(\tau)$ is defined in Eqs. (\ref{g2}).
It is seen that $W_\pi>0$, because the
$\pi$-pulse does not couple properly with the transversal degrees of freedom
excited by the furst pulse. Thus, both pulses $\CP_1$ and $\CP_\pi$
waste work.

Ultimately, the total work $W=W_1+W_\pi+W_2$ done by the three pulses
together is derived in Appendix \ref{spinechopulses} to be
\BEA
\label{w3}
W=&&\frac{\hbar\,G}{2}\,\left(1+c^{(2)}_{z,z}\,c^{(1)}_{z,z}\right)
+\hbar\,G(\tau)\,(2-c^{(2)}_{z,z}-c^{(1)}_{z,z})
-\frac{\hbar\,G(2\tau)}{2}\,\left(1+c^{(2)}_{z,z}\,c^{(1)}_{z,z}
-c^{(1)}_{z,z}-c^{(2)}_{z,z}\right)\\
&&
\label{w333}
+e^{-4\xi(\tau)+\xi(2\tau)}\,
\Re\left\{c^{(1)}_{-,z}\,c^{(2)}_{z,+}\,
\left(\,[\,2\hbar\,\dot{\xi}(\tau)-\hbar\,\dot{\xi}(2\tau)\,]
\,[\,\sin\chi_3+i\,m\cos\chi_3\,]
-\hbar\,g_3\,[\cos\chi_3-i\,m\sin\chi_3]\,
\right)\,\right\}\\
\label{stress}
&&-E\left(1+c^{(2)}_{z,z}\,c^{(1)}_{z,z}\right)+
e^{-4\xi(\tau)+\xi(2\tau)}\,
\Re\left\{c^{(1)}_{-,z}\,c^{(2)}_{z,+}\,
(\,2\,E\,\cos\chi_3-i\,\hbar\,\Omega_0\sin\chi_3\,)\,
\right\},
\label{w33}
\EEA
where 
\begin{gather}
g_3(\tau)=G-G(2\tau)
=\frac{\gamma\Gamma}{1+4\tau^2\Gamma^2},\\
\label{chi3}
\chi_3(\tau)=2F(\tau)-F(2\tau)
=\gamma\,\left[\,
\arctan(2\tau \Gamma)-2\arctan(\tau \Gamma)\,
\right],
\end{gather}
are the backreaction factors for the considered setup of pulses, and
where
\BEA
m=-
\int\d \Omega\, P(\Omega)\tanh\frac{\beta_{\rm S}\hbar\Omega}{2}<0
\EEA
is the average magnetization of the ensemble.

As compared to Eqs.~(\ref{w2}, \ref{w222}, \ref{w22}) which present
the work for two pulses, Eqs.~(\ref{w3}, \ref{w333}, \ref{w33}) are
different in several aspects. 
\begin{enumerate}

\item There are no oscillating factors $e^{i\Omega\tau}$ which after
averaging over the distribution $P(\Omega)$ would produce damping on
times ${\cal T}_2^*$. This is due to the $\pi$-pulse (\ref{pi}) in the
middle of two pulses (spin-echo setup). A simple explanation on why the
terms $\propto e^{i\Omega\tau}$ are absent is as follows.  Assume that
the interaction with the bath is absent and the spin moves under
dynamics generated by the free Hamiltonian
$\HS=\frac{\hbar\Omega}{2}\,\siz$. 
Denote by $\CE^{(0)}_t$ the
corresponding Heisenberg evolution operator: $\CE^{(0)}_t\,\hat{A}=
\exp\left[\frac{it}{\hbar}\,\HS\right]\,\hat{A}\,
\exp\left[-\frac{it}{\hbar}\,\HS\right]$. It is now seen with help of
(\ref{pi}) that the factor $e^{i\Omega\tau}$ drops out
(as if the time had been reversed):
\BEA
\CE^{(0)}_\tau\,\CP_2\,\CE^{(0)}_\tau\,\sip=
e^{i\Omega\tau}\CE^{(0)}_\tau\,\CP_2\,\sip= 
e^{i\Omega\tau}\CE^{(0)}_\tau \simin 
= e^{i\Omega\tau} e^{-i\Omega\tau} \simin = \simin.
\EEA

\item 
The decay (decoherence)
factor $e^{-4\xi(\tau)+\xi(2\tau)}$ in Eqs.~(\ref{w333},
\ref{w33}) is different from $e^{-\xi(2\tau)}$. The last decay factor 
is the one generated by the free (unpulsed) evolution during the time $2\tau$.
Only in the exponential regime
$\xi(t)\simeq t/{\cal T}_2$ we shall have
$e^{-4\xi(\tau)+\xi(2\tau)}\simeq e^{-\xi(2\tau)}$.
(Recall that the exponential regime is present
for the ohmic spectrum at long times, 
see section \ref{sec:ohm}.) For gaussian decay
$\xi(t)\simeq t^2/{\cal T}_2^2$, $e^{-\xi(2\tau)}$ predicts
sizable decay in contrast to $e^{-4\xi(\tau)+\xi(2\tau)}\simeq 1$.
This partial inhibition of decay due to $\pi$-pulse(s) is known
in NMR physics \cite{slichter} and has been recently reinterpreted as
a quantum error corection scheme \cite{viola,lidar}.

\item Now there are two independent parameters which characterize the
initial state of the ensemble of spins: $E$ and $m$. 
The work $W$ in Eqs.~(\ref{w3}, \ref{w333}, \ref{w33}) can be
expressed in the dimensionless form similar to Eq.~(\ref{dimo}):
\BEA
W=\frac{\hbar\gamma \Gamma}{2}\, w\left(
\frac{T}{\hbar\Gamma},\,\gamma, \,
\frac{\Omega_0}{\Gamma},\, \frac{T_{\rm S}}{\hbar\Gamma},\,
\frac{d}{\Gamma^2},\,
\,\tau\Gamma\right).
\EEA
It is now more convenient to account for the temperature of the spin via
$\frac{T_{\rm S}}{\hbar\Gamma}$, and there is a new dimensionless parameter
$\frac{d}{\Gamma^2}$ which quantifies the ratio of the response
time $1/\Gamma$ to ${\cal T}^*_2=1/\sqrt{d}$.
The average magnetization $m$ is expressed via 
$\frac{T_{\rm S}}{\hbar\Gamma}$ and $\frac{d}{\Gamma^2}$.

\end{enumerate}


\begin{figure}[ht] 
\includegraphics[width=7cm]{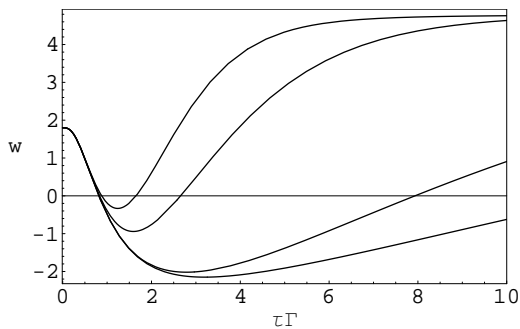}
\caption{Dimensionless work for three spin-echo pulses.
$\frac{T}{\hbar\Gamma}=10,\,5,\,1,\,0.5$ (from top to bottom), 
$\frac{T_{\rm S}}{\hbar\Gamma}=10^{3}$,
$\frac{d}{\Gamma^2}=10^{2}$,
$\gamma=0.1$,
$\frac{\Omega_0}{\Gamma}=8$. Work-extraction is poor or disappears for
smaller $\frac{d}{\Gamma^2}$ or 
$\frac{\Omega_0}{\Gamma}$, because there is
too much random thermal energy in the ensemble. 
} 
\label{f33} 
\end{figure} 

Fig.~\ref{f33} describes a scenario of work-extraction in the regime
$T_{\rm S}>T$ and for pulses
\BEA
\CP_1=\CP\left(\frac{\pi}{2};x\right),
\qquad
\CP_2=\CP\left(-\frac{\pi}{2};y\right).
\EEA
This choice of pulses amounts to substituting 
\BEA
c^{(1)}_{-,z}=i,\quad
c^{(2)}_{z,+}=\frac{1}{2},\quad
c^{(2)}_{z,z}=0,\quad c^{(1)}_{z,z}=0,
\EEA
in (\ref{w3}, \ref{w333}, \ref{w33}).

Recall that in the regime $T_{\rm S}>T$ there is a positive
contribution to the total work coming from the bath, and sizable
average frequencies $\frac{\Omega_0}{\Gamma}\geq 5$ are needed to
overcome this contribution, as seen from Fig.~\ref{f33}. This
restriction on the (average) frequency is similar to the one present
in the two-pulse work-extraction scenario for the non-disordered
ensemble of spins in the regime $T_{\rm S}>T$.

It is seen from Fig.~\ref{f33} that the initial high-temperature
ensemble of spins is strongly disordered:
$\frac{d}{\Gamma^2}=10^{2}\gg 1$.  This ratio cannot be much larger,
since there will be too much random energy in the ensemble, that is
the positive term $-E$ in Eq.~(\ref{stress}) will be too large and
cannot be compensated by potential negative terms.  Simultaneously,
the average magnetization $|m|$ will be too small.  For the same
reasons there are no interesting scenarios of work-extraction for
strongly disordered ensemble in the regime $T_{\rm S}<T$: the average
magnetization $|m|$ is too small.

\renewcommand{\thesection}{\arabic{section}}
\section{Conclusion.}
\setcounter{equation}{0}\setcounter{figure}{0} 
\renewcommand{\thesection}{\arabic{section}.}

This paper describes several related scenarios of work-extraction
based on the spin-boson model: spin-$\frac{1}{2}$ interacting with
external sources of work and coupled to a thermal bath of bosons. The
work-sources act only on the spin, since the bath is viewed as
something out of any direct access. The model has two basic
characteristic features. First, the transversal relaxations time
${\cal T}_2$ is assumed to be much shorter than the longitudinal
relaxation time ${\cal T}_1$.  This condition allows the
notion of local equilibrium, because once transversal components decay
at time ${\cal T}_2$, the spin can be described via a temperature
different from the one of the bath.  Second, the external fields are
acting in the regime of short and strong pulses. This feature makes
the analytical treatment feasible. Both these idealizations are well
known in NMR/ESR physics and related fields, and were
applied and discussed extensively in literature
\cite{slichter,nmr,abo,viola,lidar}.  It may be of interest to see in
future how precisely finite ${\cal T}_1$-times and finite
pulsing-times influence the work-extraction effect.

The work is extracted from an initial local-equilibrium state of the
spin at temperature $T_{\rm S}$ which is not equal to the temperature
$T$ of the equilibrium bath. As we recalled several times, Thomson's
formulation of the second law prohibits work-extraction via cyclic
processes from an equilibrium state of the entire system: $T=T_{\rm
S}$
\cite{thirring,bk,bassett,woron,lenard,ANthomson}. In this spirit one
would expect that work-extraction is also absent when 
external fields are acting only on the spin in a local equilibrium state
\cite{slichter,abo}. We have shown however, that this is not the case. 
It is possible to extract work in this latter setup due to the common
action of the following factors: {\it i)} backreaction of the spin to
the thermal bath; {\it ii)} generation of coherences (i.e.,
transversal components of the spin) during the work extraction
process. 

With help of the spin-echo phenomenon it is possible to extract work from
a disordered ensemble of spins having random frequencies. This
ensemble can even be strongly disordered in the sense that the
relaxation time ${\cal T}^*_2$ induced by the disorder is much smaller
than both the ${\cal T}_2$-time and the response time of the bath.

As to provide further perspectives on the obtained results,
let us discuss them in two related contextes, those of lasing
without inversion and quantum heat engines.

\subsection{Comparing with lasing without inversion.}

As we discussed in the introduction, besides the standard lasing
effect, where work is extracted from a spin having population
inversion (i.e. having a negative temperature), there are schemes of
lasing which operate with a weaker form of non-equilibrium, since they
employ three or higher-level atoms which are initially in a state with
non-zero coherences (i.e. non-zero off-diagonal elements of the
density matrix in the energy representation). There are numerous works
both theoretical and experimental, partially reviewed in
\cite{opt,olga}, showing that in such systems one can have various
scenarios of lasing without inversions in populations of atomic energy
levels. In quantum optics lasers without inversion are expected to
have several advantages over the ones with inversion.

The effects described by us also qualify as lasing without inversion (or
more precisely gain or work-extraction without inversion). There are,
however, several important differences as compared to the known
mechanisms. 
\begin{itemize}

\item
We do not require coherences present in the initial
state. Our mechanism operates starting from initial local equilibrium
state of the spin, which by itself is stable with respect to decoherence
(i.e., to both ${\cal T}_2$ and ${\cal T}^*_2$ time-scales).  It does
employ coherences however, but they are generated in the course of the
work-extraction process, which, in particular, means that all the energy
costs needed for their creation are included in the extracted work.

\item
We do not need to have three-level systems: the effect is seen
already for two-level ones. 

\item
In one of our scenarios the extracted work comes from
the bath if its temperature is higher than the initial temperature of
the spin. Due to this fact, the extracted work can be much larger than
the energy change of the spin. Thus the work extracted per cycle of 
operation can be much larger than for the standard lasing mechanism,
where it is of order of the spin's energy.

\item
The work is extracted due an initial difference between the temperature
of the spin and that of the bath. This difference can be created, e.g., 
by cooling or heating up the bath on times shorter
than the ${\cal T}_1$-time. Alternatively, one can cool or 
heat up the spin with the same restriction on the times.
The latter preparation of initially non-equilibrium
state is similar to the analogous one in the standard lasing mechanism,
except that no population inversion has to be created (i.e., no overcooling
of the spin), and the spin's temperature can be increased or decreased.

\end{itemize}

\subsection{Comparing with quantum heat engines.}

The standard thermodynamic model of a heat engine is a system (working
body) operating cyclically between two thermal baths at different
temperature and delivering work to an external source
\cite{landau,balian}.  The work pruduced during a cycle, as well as the
effciency of the production, depend on the details of the operation. The
upper bound on the effciency is given by Carnot expression, which is
system-independent (universal).  This efficiency is reached for the
Carnot cycle during very slow (slower than all the characteristic
relaxation times) and therefore reversible mode of operation
\cite{landau,balian}.  Though Carnot's cycle illustrates the best
efficiency ever attainable, it is rather poor as a model for a real
engine. This is explicitly caused by the very long duration of its
cycle: the work produced in a unit of time is very small (zero
power). This problem initiated the field of finite-time thermodynamics
which studies, in particular, how precisely the efficiency is to be
sacrified so as to reach a finite power of work \cite{ftt}.

In a similar spirit a number of researchers transferred these ideas into
quantum domain designing models for engines where the basic setup of the
classic heat engine is retained, while the working body operating
between the baths is quantum \cite{alicki,kosloff,chen1,marlan,linke}.  

Our setup for work-extraction can also viewed as model for a quantum
engine.  It is, however, of a nonstandard type since there is no working
body operating between two different-temperature systems (in our case
these are the bosonic thermal bath and the ensemble of spins). The two
systems couple directly and the work-source is acting on only one of
them.  In spite of this difference, the notion of efficiency can
be defined along the standard lines, and it is equally useful as the
standard one; in particular, it is always bound from above by the Carnot
value. We have shown that the efficiency can approach this value at the
same time as the extracted work approaches its maximum. This
is a necessary condition for a large efficiency to be
useful in practice.  Moreover, the whole process of work-extraction
takes a finite time of order of the response time of the bosonic bath,
which is actually much smaller than relaxation times of the spin. Thus,
the three desired objectives can be achieved simultaneously: maximal
work, maxmal efficiency and a large power of work.

\appendix

\section{Quantum noise generated by ohmic bath.}
\label{ohmo}

Here we discuss properties of the function:
\BEA
K(t)=\gamma \int _0^{\infty}\d \om\,\om\,\coth
(\hbar\om\beta /2)\,e^{-\om /\Gamma}\,\cos \om\,t.
\label{a1}
\EEA
In the given integration domain one can use
\BEA
\label{a13}
\coth
(\hbar\om\beta /2)=1+2\,\frac{1}{e^{\hbar\om\beta}-1}
=1+2\sum_{n=1}^{\infty}e^{-\hbar\om\beta n}
\EEA
and get from (\ref{a1}):
\BEA
K(t)= \gamma \Gamma ^2\,\frac{1-\Gamma ^2
t^2}{(1+\Gamma ^2t^2)^2}+
2\gamma \sum_{n=1}^{\infty}
\frac{(\Gamma ^{-1}+\hbar\beta n)^2-
t^2}{((\Gamma ^{-1}+\hbar\beta n)^2+t^2)^2}
\label{a3}
\EEA
With help of a standard relation:
\BEA
\sum_{n=1}^{\infty}\frac{1}{t^2+y^2(n+\kappa )^2}=
\frac{i}{2ty}\left [
\psi \left (1+\kappa -i\frac{t}{y}\right )
-\psi\left (1+\kappa +i\frac{t}{y}\right )
\right ],
\EEA
where $\psi (z)=\Gamma '(z)/\Gamma (z)$, one obtains
\BEA
\label{a2}
\sum_{n=1}^{\infty}
\frac{(\kappa +n)^2-t^2}{((\kappa +n)^2+t^2)^2}=\frac{1}{2}[
\psi '(1+\kappa -it)+\psi '(1+\kappa +it)].
\EEA
Combining (\ref{a2}) with (\ref{a3}) and $\kappa=1/(\hbar\beta\Gamma
)$ one ends up with the following formula
\BEA
K(t)= \gamma \Gamma ^2\,\frac{1-\Gamma ^2
t^2}{(1+\Gamma ^2t^2)^2}+
\frac{\gamma T^2}{\hbar^2}
\left [
\psi ' \left (1+\frac{1}{\hbar\Gamma\beta }
-i\frac{t}{\hbar\beta}\right )+
\psi ' \left (1+\frac{1}{\hbar\Gamma\beta }
+i\frac{t}{\hbar\beta}\right )
\right ]
\label{a5}
\EEA
Let us now consider separately the cases of low and high
temperatures. For $\hbar\Gamma \beta \gg 1$ one uses the known
relation 
\BEA
\label{eon}
\Gamma \left (1
-i\frac{t}{\hbar\beta}\right )
\Gamma \left (1
+i\frac{t}{\hbar\beta}\right )
=\frac{\pi t}{\hbar\beta}\,\frac{1}{\sinh [\pi t/(\hbar\beta )]}
\EEA
and obtains from (\ref{a5}):
\BEA
K(t)= \gamma \Gamma ^2\,\frac{1-\Gamma ^2
t^2}{(1+\Gamma ^2t^2)^2}+\frac{
\gamma}{t^2}
-\frac{\gamma T^2\pi^2}{\hbar^2}\,
\frac{1}{\sinh ^2[\pi t/(\hbar\beta )]}.
\EEA
For small $t$ ($t\ll 1/\Gamma$) $K(t)$ it is positive as it should be:
\BEA
K(t)=\gamma\Gamma ^2+\frac{\gamma T^2\pi^2}{3\hbar^2}.
\EEA
In contrast for $t\sim \hbar\beta\gg 1/\Gamma$ 
it becomes negative, namely the noise is anticorrelated,
\BEA
K(t)=3\gamma\,\frac{1}{\Gamma ^2t^4}
-\frac{\gamma T^2\pi^2}{\hbar^2}\,
\frac{1}{\sinh ^2[\pi t/(\hbar\beta )]}.
\label{P}
\EEA
At the end it is again correlated in the limit of very large times
$t\gg\hbar\beta$ where the first term in the r.h.s. of Eq.~(\ref{P})
dominates (this domain is shrunk for low temperatures).

In the high-temperature limit $\hbar\beta\Gamma\ll 1$ one can use in
Eq.~(\ref{a5}) the Stirling formula:
\BEA
\psi '(z)=\frac{1}{z}+\frac{1}{2z^2}+...,\qquad z\ge 1
\EEA
and then the quasiclassical limit for the quantum noise reads
(after some more simplifications):
\BEA
K(t)= \gamma \Gamma ^2\,\frac{1-\Gamma ^2
t^2}{(1+\Gamma ^2t^2)^2}+
\frac{2\gamma T\Gamma}{\hbar}\,
\frac{1}{1+t^2\Gamma ^2}.
\label{punic}
\EEA
In the purely classical limit the first term in the r.h.s. can be
neglected and we return (for $t\Gamma \gg 1$) to the classical white
noise with the strength $2\pi\gamma T$.

Finally in the context of Eq.~(\ref{a5}) we notice the following
useful relations:
\BEA
\dot \xi(t)=\int _0^t\d t'\,K(t')= 
\gamma\Gamma \left(\,\frac{t\Gamma
}{1+\Gamma ^2t^2}+
\frac{iT}{\hbar\Gamma}
\left [
\psi  \left (1+\frac{1}{\hbar\Gamma\beta }
-i\frac{t}{\hbar\beta}\right )-
\psi  \left (1+\frac{1}{\hbar\Gamma\beta }
+i\frac{t}{\hbar\beta}\right )
\right ]\,\right),
\label{a20}
\EEA
\BEA
\xi(t)=
\int_0^t\d t'\int _0^{t'}\d t''\,K(t'')
= \gamma \,
\ln \left [\frac{\Gamma ^2\left(
1+\frac{1}{\hbar\Gamma\beta }
\right)\sqrt{1+t^2\Gamma ^2}}
{\Gamma\left (1+\frac{1}{\hbar\Gamma\beta }
-i\frac{t}{\hbar\beta}\right )
\Gamma \left (1+\frac{1}{\hbar\Gamma\beta }
+i\frac{t}{\hbar\beta}\right )}
\right],
\label{a30}
\EEA
which are used in the main text.

\section{}
\label{acooling}

Here we shortly outline how the two-temperature state (\ref{ka})
can be prepared starting from the overall equlibrium state
\BEA
\rho(0)=\frac{1}{Z}\exp\left[
-\beta_{\rm S} \HS-\beta_{\rm S} (\HI+\HB)
\right],\qquad Z=\tr \,e^{-\beta_{\rm S} \HS-\beta_{\rm S} (\HI+\HB)},
\label{choban}
\EEA
which has equal temperatures of the spin and the bath.

Assume that the bath was subjected to another much larger thermal bath
(superbath) at temperature $T$ different from $T_{\rm S}$, so that 
the total Hamiltonian of the spin, bath and superbath reads:
\BEA 
\HH_{\rm total}=\HH+\HH_{\rm sup}, 
\EEA 
where the latter operator $\HH_{\rm sup}$ characterizes the {\it weak}
interaction of
the bath with the superbath and contains also self-Hamiltonian of the
superbath. Thus
\BEA
[\HS,\HH_{\rm sup}]=0.
\label{mr}
\EEA

Now the statement of this appendix is that under the action of the superbath
at temperature $T$, the common state of the spin and the bath will relax
to the state Eq. (\ref{prom} or \ref{ka}) with different temperatures 
for the spin and the
bath. The reason is that due to Eq. (\ref{mr}), $\siz$ is conserved
during the whole evolution generated by the superbath, so that 
$\siz$ does not relax and keeps its value given by 
Eq. (\ref{choban})~\footnote{
In a more realistic situation, where ${\cal T}_1$-time is kept finite,
the relaxation time of the bath under action of a superbath has to be
much smaller that ${\cal T}_1$, so as to create the temperature
difference between the spin and the bath.}
In contrast, the variables of the bath |including $\X$| 
do not have such a protection, so they relax under influence of the 
superbath. Let us now substantiate this statement.

Because $[\HS,\HH]=0$, the initial equilibrium state 
$\rho(0)$ of the spin and the bath can be represented as
\BEA
\rho(0)=\sum_{j=\pm 1}p_{jj}\,\rho_{jj}(0)\,|j\rangle\langle j|,
\label{kr}
\EEA
where 
\BEA
p_{jj}=\frac{e^{-j\beta_{\rm S}\eps/2}}{2\cosh(\beta_{\rm S}\eps/2)},
\qquad j=\pm 1,
\EEA
are probabilities for the spin to be up or down respectively, 
$|j\rangle$ is the eigenstate of $\HS=\frac{\eps}{2}\siz$ with eigenvalue
$j=\pm 1$, and where
\BEA
\rho_{jj}(0)=
\frac{1}{Z_{j}}\exp\left[
-\beta_S \left(\frac{j}{2}\X+\HB\right)
\right],\qquad Z_j=\tr_{\rm B} \,e^{-\beta_S (\frac{j}{2}\X+\HB)},
\qquad j=\pm 1,
\EEA
are conditional states of the bath.

The total initial state of the spin, bath and superbath thus reads:
\BEA
\rho_{\rm total}(0) =\sum_{j=\pm 1}p_{jj}\,\rho_{\rm
sup}(0)\otimes\rho_{jj}(0)\,|j\rangle\langle j|, 
\EEA 
where $\rho_{\rm sup}(0)$ is the initial equilibrium state of the
superbath.  Note that due to weak coupling between the bath and
superbath, their initial states can be assumed to be factorized.

As follows from (\ref{mr}, \ref{kr}), 
the time-dependent state of the total system consisting
of spin, bath and superbath can be presented as
\BEA
\rho_{\rm total}(t)
=\sum_{j=\pm 1}p_{jj}\,\Omega_{jj}(t)\,|j\rangle\langle j|,
\EEA
where $\Omega_{jj}(t)$ |the conditional joint state of the bath and superbath |
satisfies the von Neumann equation
\BEA
\i\hbar\dot{\Omega}_{jj}=[\frac{j}{2}\X+\HB+\HH_{\rm sup},
\Omega_{jj}],
\EEA
with the initial condition:
\BEA
\Omega_{jj}(0)=\rho_{\rm sup}(0)\otimes\rho_{jj}(0).
\EEA

Thus, $\Omega_{jj}$ moves according to the Hamiltonian 
$\frac{j}{2}\X+\HB+\HH_{\rm sup}$. It is now clear that
in the weak coupling limit of the bath-superbath interaction
the marginal conditional state $\tr_{\rm sup}\Omega_{jj}(t)$
will | for sufficiently long times $t$ | relax to Gibbs distribution 
at temperature $T$ (equal to the one of the superbath) and with
Hamiltonian $\frac{j}{2}\X+\HB$. Thus the (unconditional) marginal state
of the spin and the bath will indeed relax to
\BEA
\rho\propto \exp\left[
-\beta_S \HS-\beta (\HI+\HB)
\right].
\label{prom}
\EEA

\section{}
\label{maugli}

Here we explain in detail why the initial conditions (\ref{fedor}, \ref{fedor1})
and (\ref{ka}) are equivalent.

One can write the full Hamiltonian $\HH$ defined in (\ref{ham}) as
\BEA
\label{hamo}
\HH=\sum _k\hbar\om _k\,
\left(\ha^{\dagger}_k+\frac{g_k\siz}{2\om_k}
\right)\left(
\ha _k+\frac{g_k\siz}{2\om_k}\right)+\frac{\eps}{2}\,\siz-
\sum _k\frac{\hbar g^2_k}{4\om_k},
\EEA
and diagonalize it via a unitary operator:
\BEA
\U=\exp\left[
\sum _k \frac{g_k\siz}{2\om_k}
(\ha _k^{\dagger}-\ha _k)
\right],\qquad
\U\,\ha_k\,\U^\dagger=
\ha_k-\frac{g_k\siz}{2\om_k}, \qquad
\U\,\siz\,\U^\dagger=\siz.
\EEA

Thus the operators
\BEA
\label{esh}
\hb_k=\ha_k+\frac{g_k\siz}{2\om_k},\qquad
[\hb_k,\hb_l^\dagger]=\delta_{kl}
\EEA
are distributed |over the initial state (\ref{ka})| independently from
the operators of spin.  Moreover, as follows from (\ref{hamo},
\ref{ka}), the operators $\hb_k$ have on the state (\ref{ka}) exactly
the same statistics (i.e., the same correlators) as the corresponding
operators $\ha_k$ on the factorized state (\ref{fedor}).

Now note that for the initial condition (\ref{ka}), $\siz(0)$ and the
quantum noise operator $\e(t)$ are in general not independent
variables, in contrast to the case of the factorized initial condition
(\ref{fedor}, \ref{fedor1}). However, for $t\gg 1/\Gamma$ they do become
independent:
\BEA
\label{kov}
\e(t)=\e_b(t)+\siz (G(t)-G),\qquad
\e_b(t)\equiv
\sum_kg_k
[\hb_k^{\dagger}(0)e^{i\om _kt}+\hb_k(0)e^{-i\om _kt}],
\EEA
where 
\BEA
\label{shopen}
G\equiv \sum_k\frac{g_k^2}{\om_k}
\EEA
is the limit of $G(t)$ for $t\gg 1/\Gamma$.  Taking the latter
limit in (\ref{kov}), one gets $\e(t)$ is equal to $\e_b(t)$ and
is thus independent of $\siz$. Recalling that $\e_b(t)$ has on the
state (\ref{ka}) the same statistics as $\e(t)$ on the factorized
state (\ref{fedor}), finishes the argument: the equivalence holds 
for times larger than the bath response time ($1/\Gamma$ for the 
ohmic situation).

Note that the thermodynamical limit
for the bath is essential for this conclusion.  Otherwise, $G(t)$ would
qbe a finite sum of cosines, and would not converge to $G$.

\section{Derivation of Eqs.~(\ref{fa1}, \ref{fa2}).}
\label{acarnot}

Assume that the initial state of the spin and bath is:
\BEA
\rho(0)=\frac{1}{Z}\exp\left[
-\beta_S \HS-\beta (\HI+\HB)
\right],\qquad Z=\tr \,e^{-\beta_S \HS-\beta (\HI+\HB)},
\label{do}
\EEA
with different temperatures for the spin and the bath.

An external field $\hat{V}(t)$ is acting on the system, 
\BEA
\HH(t)=\HH+\hat{V}(t)
\EEA
such
that it is zero both initially and at the moment $t=\tau$:
\BEA
\hat{V}(\tau)=\hat{V}(0)=0.
\EEA
This condition defines {\it cyclic process}. 
The total work which was done on this system reads:
\BEA
W=\Delta H_{\rm S}+\Delta H_{\rm I}+\Delta H_{\rm B},
\label{mi}
\EEA
where
\BEA
\HH_{\rm k}=\tr\left (\HH_{k}\, [\,
\rho(\tau)-\rho(0)\,]
\right), \qquad {\rm k}={\rm S}, {\rm I}, {\rm B},
\EEA
are the changes of the corresponding energies, and where
$\rho(\tau)$ is the overall density matrix at time $\tau$.

Recall that the relative entropy (see, e.g. \cite{petr}):
\BEA
S[\rho ||\sigma]\equiv\tr (\rho\ln\rho-\rho\ln\sigma)\geq 0,
\EEA
is non-negative for any density matrices $\rho$ and $\sigma$.
One now uses:
\BEA
S[\rho(\tau) ||\rho(0)]&&=\tr (\,\rho(\tau)\ln\rho(\tau)-\rho(\tau)
\ln\rho(0)\,)=\tr (\,\rho(0)\ln\rho(0)-\rho(\tau)
\ln\rho(0)\,)\nonumber\\
&&=\beta_S\Delta H_{\rm S}+\beta
(\Delta H_{\rm I}+\Delta H_{\rm B})\geq 0,
\label{re}
\EEA
where we used (\ref{do}) and $\tr\rho(\tau)\ln\rho(\tau)=
\tr\rho(0)\ln\rho(0)$ is due to the unitarity of the overall dynamics
generated by the time-dependent Hamiltonian $\HH(t)$.

Combining (\ref{re}) with (\ref{mi}) one gets Eqs.(\ref{fa1}, \ref{fa2})
\BEA
\label{shu}
W\geq \left(1-\frac{T}{T_S}\right)\,\Delta H_{\rm S},
\qquad
W\geq \left(1-\frac{T_S}{T}\right)\,(\Delta H_{\rm I}+\Delta H_{\rm B}).
\EEA

Finally note that would we use the initial conditions
\BEA
\label{stu}
\rho (0)=\rho _{\rm S}(0)\otimes \rho _{\rm B}(0)=
\frac{1}{\tr\,e^{-\beta \HS}}e^{-\beta \HS}\otimes
\frac{1}{\tr\,e^{-\beta \HB}}e^{-\beta \HB}
\EEA
we would not be able to conclude from the above derivation that
the efficiency is limited by the Carnot value. 
Indeed, instead of Eqs.~(\ref{shu}) one has, respectively:
\BEA
\beta_S\Delta H_{\rm S}
+\beta\Delta H_{\rm B}\geq 0,
\qquad
W\geq \Delta H_{\rm I}+
\left(1-\frac{T}{T_S}\right)\,\Delta H_{\rm S}.
\EEA
The latter inequality is not informative with respect to Carnot's bound, 
since it cannot and should not in general be excluded that 
$\Delta H_{\rm I}$ is sizeable.

However, for the model studied in the present paper, the equivalence
of the initial conditions (\ref{do}) and (\ref{stu}) is known from
other places.

Let us emphasize the main points by which the present derivation differs
from the standard textbook one:

\begin{itemize}

\item No postulates were used: the whole derivation is based on the
quantum-mechanical equations of motion and certain assumptions on the
initial conditions.

\item It was not assumed that the interaction between the system and
the bath is small, a restrictive assumptions which need not be
satisfied in reality. 

\item The fact of using the initial conditions in Eq. (\ref{do}) 
is important 
in the present derivation, though presumably Carnot's bound is valid in
certain more general cases, such as, in our case, factorized 
initial conditions from Eq. (\ref{stu}).

\end{itemize}

\section{Some correlation functions.}

In this appendix and in the following ones we study veraious dynamical
aspects of the model 
defined by Eqs.~(\ref{ham}, \ref{hamspin},
\ref{hambath},\ref{hamint}). 
The initial conditions are given by (\ref{fedor},
\ref{fedor1}).
Hereafter $\langle ...\rangle$ means averaging over this initial
condition.

Let us define some correlation functions.

\paragraph{}

For 

\BEA
t_3\geq t_2\geq t_1,
\label{duda}
\EEA
and recalling definitions (\ref{eta=}, \ref{pi=})
one derives using Wick's theorem
in the same way as when deriving (\ref{ugar}):
\BEA
\left\langle\e (t_3)\,\,\pipm(t_1,t_2)\,\right\rangle
&&=\pm\sum _{k=0}^{\infty}\,\frac{i\,(-1)^{k}}{(2k+1)!}
\int_{t_1}^{t_2}...\int_{t_1}^{t_2}\, \d s_1...\d s_{2k+1}\left \langle
{\cal T}\, [\,\e (t_3)\e (s_1)...\e (s_{2k+1})\,]\right \rangle\nn \\
\label{d111}
&&=\pm i\,\int_{t_1}^{t_2}\,\d s\,K_{\cal T}(t_3-s)\,\,
\exp \left [-\frac{1}{2}\,\int_{t_1}^{t_2}
\int_{t_1}^{t_2}\,\d s_1\,\d s_2\,
K_{\cal T}(s_1-s_2)\right ]\\
&&=\pm[\,i\dot\xi(t_3-t_1)-i\dot\xi(t_3-t_2)+G(t_3-t_1)
-G(t_3-t_2)\,]\,
e^{-\xi(t_2-t_1)+i\,F(t_2-t_1)}.
\label{ugar-1}
\EEA
where for deriving the last line we used the definition
of $K_{\cal T}(t)$:
\BEA 
K_{\cal T}(t)=K(t)-i\dot G(t)
=\ddot\xi(t)-i\ddot F(t).
\EEA

Note that for $t_3=t_2$ we can derive (\ref{ugar-1}) in a simpler way
by employing (\ref{ugar}) and 
\BEA
\left\langle\e (t_2)\,\pipm(t_1,t_2)\,\right\rangle
=\mp i\,\partial_{t_2}
\left\langle\pipm(t_1,t_2)\,\right\rangle.
\label{morda}
\EEA

\paragraph{}

A correlation function $\left\langle\pipm(t_1,t_2)\,\e
(t_3)\right\rangle$ under the same condition (\ref{duda}) is studied
similarly to (\ref{ugar-1}), the only difference being that the
time-ordered correlation function $K_{\cal T}(t_3-s)$ in (\ref{d111})
is substituted by the analogous time-antiordered one
(time-antiordering comes due to (\ref{duda}))
\BEA
K_{\cal A}(t_3-s)=K^*_{\cal T}(t_3-s).
\EEA
These two functions are related by complex conjugation, as seen from
(\ref{02}). Thus,
\BEA
\left\langle\pipm(t_1,t_2)\,\e (t_3)\right\rangle=
\pm[\,i\,\dot\xi(t_3-t_1)-i\,\dot\xi(t_3-t_2)-G(t_3-t_1)
+G(t_3-t_2)\,]\,
e^{-\xi(t_2-t_1)+i\,F(t_2-t_1)}.
\label{ugar-11}
\label{e5}
\EEA
As compared to (\ref{ugar-1}), the sign of $G$-factors is seen to change.

\paragraph{}

A correlation function between two $\hat{\Pi}$-factors for
\BEA
t_4\geq t_3\geq t_2\geq t_1,
\EEA
is worked out as follows. First one notes
\BEA
\left\langle\pipm(t_3,t_4)\,\pimp(t_1,t_2)\right\rangle=
\left\langle{\cal T}
\exp\left[\pm i\,\int_{t_1}^{t_4}\d s\,\phi(s)\,\e(s)
\right]
\right\rangle,
\EEA
where
\BEA
\phi(s)&&=-1, \quad t_1\leq s\leq t_2,\nonumber\\
       &&=0, \quad t_2\leq s\leq t_3,\nonumber\\
       &&=1, \quad t_3\leq s\leq t_4.
\EEA

One gets

\BEA
&&\left\langle\pipm(t_3,t_4)\,\pimp(t_1,t_2)\right\rangle=
\exp \left [-\frac{1}{2}\,\int_{t_1}^{t_2}
\int_{t_1}^{t_2}\,\d s_1\,\d s_2\,
K_{\cal T}(s_1-s_2)
-\frac{1}{2}\,\int_{t_3}^{t_4}
\int_{t_3}^{t_4}\,\d s_1\,\d s_2\,
K_{\cal T}(s_1-s_2)\right.\nonumber\\
&&~~~~~~~~~~~~~~~~~~~~~~~~~~~~~~
\left.+\int_{t_1}^{t_2}
\int_{t_3}^{t_4}\,\d s_1\,\d s_2\,
K_{\cal T}(s_1-s_2)
\right ]\nonumber\\
&&=\exp \left [
-\xi(t_2-t_1)-\xi(t_4-t_3)-\xi(t_4-t_2)+\xi(t_4-t_1)
+\xi(t_3-t_2)-\xi(t_3-t_1)
\right ]\nonumber\\
&&\times\exp \left [
iF(t_2-t_1)+iF(t_4-t_3)+iF(t_4-t_2)-iF(t_4-t_1)
-iF(t_3-t_2)+iF(t_3-t_1)
\right ].
\label{e9}
\EEA

\section{Evolution of the quantum noise and 
$\hat{\Pi}$-factors under Heisenberg dynamics.}
\label{usefulHeisenberg}

Note from (\ref{gegel},
\ref{eta=}) how the quantum noise and
$\hat{\Pi}$-factors evolve under Heisenberg dynamics:
\BEA
\label{ho1}
&&\CE_{t}\,\e(\tau)\equiv
e^{i\HH t/\hbar}\,\e(\tau)\,e^{-i\HH t/\hbar}=\e
(t+\tau)+\si_z\,[\,G(\tau)-G(t+\tau)],\\
\label{ho2}
&&\CE_{t}\,\hat{\Pi}_{\pm}(t_1,t_2)\equiv
e^{i\HH t/\hbar}\,\hat{\Pi}_{\pm}(t_1,t_2)\,e^{-i\HH t/\hbar}=
{\cal T}\exp\left[\pm i\,\int_{t_1}^{t_2}\d s \sum_kg_k
(\ha_k(t)e^{-i\omega_k s}+\ha^\dagger_k(t)e^{i\omega_k s})\right]\nonumber\\
&&=\hat{\Pi}_{\pm}(t+t_1,t+t_2)\,\exp\left(\,\pm i\,\si_z\,\left[\,
F(t_2)-F(t_1)+F(t+t_1)-F(t+t_2)\,\right]\,\right).
\label{karas}
\EEA
When deriving these equations we used $[\e(\tau+t),\si_z$]=0. Recall
that $\siz$ is conserved under evolution generated by $\HH$: 
$\CE_t\,\siz=\siz$.

\section{Derivations for two pulses.}
\label{twopulses}

The work done by the second pulse is defined as 
\BEA
\oneh W_2&&=
\oneh \left\langle
\CP_2\,(\HS+\HI)-(\HS+\HI)
\right\rangle_{t+\tau}\nonumber\\
&&=\frac{\Omega}{2}\left\langle
\CP_2\siz-\siz
\right\rangle_{t+\tau}+
\frac{1}{2}\left\langle\left(
\CP_2\siz-\siz\right)\X
\right\rangle_{t+\tau}\\
&&=\frac{\Omega}{2}\,
(c^{(2)}_{z,z}-1)\left\langle
\siz\right\rangle_{t+\tau}\,
+\frac{1}{2}\,
(c^{(2)}_{z,z}-1)\left\langle
\siz\,\X\right\rangle_{t+\tau}\,
+\Omega\,\Re\left\{c^{(2)}_{z,+}
\langle\sip\rangle_{t+\tau}\right\}
+\Re\left\{c^{(2)}_{z,+}
\,\left\langle
\sip\,\X\right
\rangle_{t+\tau}
\right\},\nonumber\\
&&
\label{barunak}
\EEA
where the averages are taken at the time $t+\tau$ immediately
before the second pulse, and where we used the definition 
(\ref{parametrization}) of the parametrization coefficients.
For clarity we recall that definition here
\BEA
\label{parametrization1}
\CP_k\,\hat{\sigma}_a
=\sum_{b=\pm,\,z}c^{(k)}_{a,b}\,\hat{\sigma}_b,
\qquad 
a=\pm,\,z, \qquad k=1,2.
\EEA

In order to calculate $W_2$ we thus have to determine
$\X(t+\tau)$, $\sip(t+\tau)$, and $\siz(t+\tau)$.
Recall from Eq.~(\ref{heisenbergpulse}) that, e.g.,
\BEA
\X(t+\tau)=
\CE_t\,\CP_1\,\CE_\tau\,\X,
\EEA
where $\CE_t$ is the free evolution (super)operator defined in
Eq.~(\ref{hhf}). One infers from (\ref{gegel}, \ref{kant}) 
\BEA
\label{g3.1}
\CE_\tau\,\X=\e(\tau)-\siz G(\tau),
\EEA\BEA
\CE_t\,\e(\tau)=\e(t+\tau)-\siz [ G(\tau)-G(t+\tau)],
\label{g3.2}
\EEA

and then

\BEA
\label{g4}
\X(t+\tau)\equiv \CE_t\,\CP_1\,\CE_\tau\,\X
=\e(t+\tau)+[\,G(\tau)-G(t+\tau)\,]\,\siz-G(\tau)\,\CE_t\,\CP_1\,\siz
\EEA

The formula for $\siz(t+\tau)$ is more straightforward:
\BEA
\siz(t+\tau)=\CE_t\,\CP_1\,\siz,
\EEA
\BEA
\label{utkanos}
\lel\siz(t+\tau)\rir=
c^{(1)}_{z,z} \lel\siz\rir,
\label{g6}
\EEA
where we noted that in 
\BEA
\label{gg}
\lel\CE_t\,\CP_1\,\si_k  \rir=\sum_{n=\pm,z}
c^{(1)}_{k,n}\lel\CE_t\,\hat{\sigma}_n  \rir=
c^{(1)}_{k,z}\lel\siz\rir,\qquad k=\pm, z,
\EEA
only one term 
contributes, 
since $\langle\si_{\pm}\rangle=\langle\, e^{i\chi\siz}
\si_{\pm}\rangle=0$ due to the initial conditions (\ref{fedor},
\ref{fedor1}).

In the same way one calculates
\BEA
\langle\siz\,\X\rangle_{t+\tau}=
[G(\tau)-G(t+\tau)]c^{(1)}_{z,z}-G(\tau),
\EEA
\BEA
\label{tup}
\label{g8}
\sip(t+\tau)\equiv \CE_t\,\CP_1\,\CE_\tau\,\sip=
e^{-iF(\tau)+i\Omega \tau}\,\CE_t\,\CP_1\piplus(0,\tau)\sip=
e^{-iF(\tau)+i\Omega \tau+i\chi\siz}\piplus(t,t+\tau)\,
\CE_t\,\CP_1\,\sip,
\label{shesh}
\EEA
where we used (\ref{karas}), $[\siz,\pipm]=0$, and where by definition
(from (\ref{karas})):
\BEA
\chi(\tau,t)=\int_0^\tau\d
s\,[\,G(s)-G(t+s)\,]=F(t)+F(\tau)-F(t+\tau).
\EEA

Now let us recall Eq.~(\ref{ugar}):
\BEA
\label{ugarugar}
\left\langle
\,\pipm(t_1,t_2)
\,\right\rangle=
\exp[-\xi(t_2-t_1)+i\,F(t_2-t_1)],
\EEA
because it is used in averaging the RHS of Eq.~(\ref{tup}):
\BEA
\langle\sip(t+\tau)\rangle=c^{(1)}_{+,z}\,e^{i\omega \tau-\xi(\tau)}
\,\langle e^{i\chi\siz}\,\siz\rangle,
\EEA
where we additionally
employed the reasoning which led us to (\ref{gg}).

The last term we have to calculate is 
$\left\langle\sip\,\X\right\rangle_{t+\tau}$.
Directly multiplying (\ref{g4}) and (\ref{shesh})
one gets
\BEA
\left\langle\sip\,\X\right\rangle_{t+\tau}=
\label{bobo1}
e^{-iF(\tau)+i\Omega \tau+i\chi\siz}\piplus(t,t+\tau)\,
\CE_t\,\CP_1\,\sip\,\e(t+\tau)\\
\label{bobo2}
+[\,G(\tau)-G(t+\tau)\,]\,
e^{-iF(\tau)+i\Omega \tau+i\chi\siz}\piplus(t,t+\tau)\,
\CE_t\{\,(\CP_1\,\sip)\,\siz\,\}\\
-G(\tau)\,
e^{-iF(\tau)+i\Omega \tau+i\chi\siz}\piplus(t,t+\tau)\,
\CE_t\{\CP_1\,\sip\siz\}.
\label{bobo3}
\EEA
Following to (\ref{parametrization1}) we now expand $\CP_1\,\sip$ in
(\ref{bobo2}--\ref{bobo3}). With the same reasoning as for (\ref{gg}),
we need to keep in these expansions only terms propotional to
$c^{(1)}_{+,z}$ (since $\langle\sipm\rangle=0$ according to the
initial conditions (\ref{fedor},
\ref{fedor1})).
After further simplifications with help of (\ref{e5})
we obtain
\BEA
\left\langle
\sip\,\X\right
\rangle_{t+\tau}=c^{(1)}_{+,z}\,e^{i\Omega\tau-\xi(\tau)}\,
\left(
i\dot{\xi}(\tau)\,\left\langle e^{i\chi\siz}\siz\right\rangle
+[\,G(\tau)-G(t+\tau)\,]\,\left\langle e^{i\chi\siz}\right\rangle
\right).
\EEA

The final formula for the work reads:
\BEA 
\oneh W_2=\half\,
(1-c^{(1)}_{z,z})(1-c^{(2)}_{z,z})\,G(\tau)
+\half\,(1-c^{(1)}_{z,z}\,c^{(2)}_{z,z})\,G(t+\tau)       
+\Omega\Re\left\{
c^{(2)}_{z,+}\,\langle
\sip
\rangle_{t+\tau}
\right\}+\Re\left\{
c^{(2)}_{z,+}\,\left\langle
\sip\,\X\right
\rangle_{t+\tau}
\right\}.
\label{gusak}
\EEA

Note that in the limit $t\to\infty$ (which means $t\gg 1/\Gamma$), one
has $G(t+\tau)\to G$, where $G$ is defined in (\ref{shopen}).

Eq.~(\ref{gusak}) can be put into dimensionless form as anounced by
(\ref{dimo}).  To this end note from (\ref{a20}, \ref{a30}) that
$\xi(\tau)$ and $\frac{1}{\Gamma}\dot{\xi}(\tau)$ can be expressed via
dimensionless quantities $\tau\Gamma$, $\gamma$ and $T/(\hbar\Gamma)$.
In the same way we note from (\ref{castro1}, \ref{castro111}) that
$\frac{1}{\Gamma}G(\tau)$ and $F(\tau)$ are expressed via $\gamma$ and
$\tau\Gamma$.

\section{Derivations for three pulses (spin-echo setup).}
\label{spinechopulses}

Now we consider three pulses, $\CP_1$, $\CP_\pi$ and $\CP_2$ which are
applied, respectively, at times $t$, $t+\tau$ and $t+2\tau$. The
pulses $\CP_1$ and $\CP_2$ are kept arbitrary, while $\CP_\pi$ is the
$\pi$-pulse defined by Eq. (\ref{pi}).

The work done for the pulse $\CP_2$ is defined by the same formula
(\ref{barunak}), where now all the averages are taken at the time
$t+2\tau$ immediately before the application of $\CP_2$. Our
calculations in the following will be relatively brief, since in
essence they follow to the pattern of calculations in the previous
appendix.

For $\sip(t+2\tau)$ we get 
\BEA
\label{tuptup}
&&\sip(t+2\tau)\equiv 
\CE_t\,\CP_1\,\CE_\tau\,\CP_\pi\,\CE_\tau\, \sip=
e^{-iF(\tau)+i\omega \tau}\,\CE_t\,\CP_1\,\CE_\tau\,
\piplus(0,\tau)\simin\nonumber\\
&&=e^{-4iF(\tau)+iF(2\tau)}\,\CE_t\,\CP_1\,\piplus(\tau,2\tau)\,
\piminus(0,\tau)\,\simin=
e^{-4iF(\tau)+iF(2\tau)}\,\CE_t\left\{\,\piplus(\tau,2\tau)\,
\piminus(0,\tau)\right\}\,\CE_t\left\{\CP_1\,\simin\right\}
\nonumber\\
&&=e^{-4iF(\tau)+iF(2\tau)}
~\piplus(t+\tau,t+2\tau)\,\piminus(t,t+\tau)\,
e^{-i\chi_3\siz}
\,\CE_t\left\{\CP_1\,\simin\right\},
\EEA
where we we used (\ref{karas}) and defined
\BEA
\chi_3(\tau,t)=2F(\tau)
-F(2\tau)-2 F(t+\tau)+ F(t)+F(t+2\tau).
\EEA
Taking in this equation the limit $t\gg 1/\Gamma$ and using
(\ref{castro111}) we return to the quantity $\chi_3(\tau)$ as defined
by (\ref{chi3}).

With help of (\ref{e9}) and the reasoning of (\ref{gg}) one has
\BEA
\lel\sip(t+2\tau)\rir=
c^{(1)}_{-,z}\,e^{-4\xi(\tau)+\xi(2\tau)}\,
\left\langle e^{-i\chi_3\siz}\,\siz\right\rangle
\EEA

In the same way as for (\ref{tuptup}) we have
\BEA
\siz(t+2\tau)\equiv
\CE_t\,\CP_1\,\CE_\tau\,\CP_\pi\,\CE_\tau\, \siz
=-\CE_t\,\CP_1\,\siz,
\EEA
while applying (\ref{g3.1}, \ref{g3.2}) one derives:
\BEA
\X(t+2\tau)\equiv 
\CE_t\,\CP_1\,\CE_\tau\,\CP_\pi\,\CE_\tau\, \X
=\e(t+2\tau)+[\,2G(\tau)-G(2\tau)\,]\,
\CE_t\,\CP_1\,\siz+
[\,G(2\tau)-G(t+2\tau)\,]\,\siz.
\EEA

The only non-trivial relation in calculating
$\left\langle\sip\,\X\right\rangle_{t+2\tau}$
is 
\BEA \left\langle
~\piplus(t+\tau,t+2\tau)\,\piminus(t,t+\tau)\,
\eta(t+2\tau)
\right\rangle
=[\,
2i\dot{\xi}(\tau)-i\dot{\xi}(2\tau)
-2G(\tau)+G(2\tau)
\,]\,\,e^{-4\xi(\tau)+\xi(2\tau)},
\EEA
which is obtained in the same way as (\ref{e5}, \ref{e9}).  The
easiest way to check this relation is to follow to the derivation of
(\ref{morda}), that is, to differentiate (\ref{e9}) over $t_4$, to put
$t_4=t+2\tau$, $t_3=t_2=t+\tau$, $t_1=t$, and then to change the sign
of all $G$-factors in the final expression.

If the reader has followed us so long, he/she can continue alone, since
the remaining calculations are fairly straightforward.

\end{document}